\documentclass{aa}
\usepackage{graphics}

\def\cm2{\,{\rm cm^{-2}}}

\def\13co{\,{\rm ^{13}CO}}
\def\h2{\,{\rm H_{2}}}
\def\aua{{\rm A\&A} }

\def\apj{{\rm ApJ} }
\def\aj{{\rm AJ} }
\def\apjs{{\rm ApJS} }
\def\apjl{{\rm ApJL} }

\def\mnras{{\rm MNRAS} }

\begin{document}

\title{Sub-millimeter to centimeter excess emission from the Magellanic Clouds. }
 
   \subtitle{I. Global spectral energy distribution.}
 
\author{F.P. Israel \inst{1},
        W.F. Wall \inst{2}	
        D. Raban \inst{1},
        W.T. Reach \inst{3},
        C. Bot \inst{4},
        J.B.R. Oonk \inst{1},
        N. Ysard \inst{5} 
and     J.P.Bernard \inst{6}   }
 
   \offprints{F.P. Israel}
 
  \institute{ Sterrewacht Leiden, Leiden University, P.O. Box 9513,
    2300 RA Leiden, The Netherlands 
       \and Instituto Nacional de Astrof\'isica, \'Optica, y
       Electr\'onica, Apdo. Postal 51 y 216, Puebla, Pue., M\'exico;
       \and Spitzer Science Center, California Institute of
       Technology, Pasadena, CA, USA 
       \and UMR7550, Observatoire Astronomique de Strasbourg,
       Universit\'e Louis Pasteur, F-67000 Strasbourg, France 
       \and Department of Physics, P.O. Box 64, FIN-00014 University 
       of Helsinki, Finland
       \and Universit\'e de Toulouse, UPS, CESR, F-31028 Toulouse,
       France}
 
\authorrunning{F.P. Israel et al.}

\date{Received ????; accepted ????}
 
\abstract{} {Our goal is to determine and study the global emission from 
  the Magellanic Clouds over the full radio to ultraviolet spectral range.}
  {We have selected from the literature those flux densities that include 
  the entire LMC and SMC respectively, and we have complemented these with 
  maps extracted from the WMAP and COBE databases covering the missing 
  the 23--90 GHz (13--3.2 mm) and the poorly sampled  1.25--250 THz 
  (240--1.25 $\mu$m) spectral ranges in order to reconstruct the global 
  SEDs of the Magellanic Clouds over eight decades in frequency or 
  wavelength.}  
  {A major result is the discovery of a pronounced excess of emission from 
  the Magellanic Clouds at millimeter and sub-millimeter wavelengths.  We 
  also confirm global mid-infrared (12$\mu$m) emission suppression, and 
  determine accurate thermal radio fluxes and very low global extinctions 
  for both LMC and SMC, the latter being the most extreme in all these 
  respects.}  
  {These and other dust properties such as the far-UV extinction curve 
  appear to be correlated with (low) metallicity. Possible explanations 
  are briefly considered. As long as the nature of the excess emission is 
  unknown, the total dust masses and gas-to-dust ratios of the Magellanic 
  Clouds cannot reliably be determined}
         {

\titlerunning{(Sub)millimeter excess in LMC and SMC}

\maketitle

\section{Introduction}

%Table 1  WMAP Data
\begin{table}
\caption[]{Global Emission from LMC and SMC.}
\begin{center}
\begin{tabular}{cccc}
\hline
\noalign{\smallskip}
Freq. & Wavel. & \multicolumn{2}{c}{Flux Densities} \\
 $\nu$    & $\lambda$  & LMC  & SMC \\
\noalign{\smallskip}
\multicolumn{4}{c}{WMAP}\\
\noalign{\smallskip}     
\hline
\noalign{\smallskip}     
\hline
\noalign{\smallskip}
(GHz)     & (mm)       &\multicolumn{2}{c}{Jy}\\
\noalign{\smallskip}     
\hline
\noalign{\smallskip}
 23  &    13    & 160$\pm$25 & 17$\pm$3  \\
 31  &     9.8  & 159$\pm$25 & 20$\pm$4  \\
 41  &     7.3  & 165$\pm$25 & 24$\pm$5  \\
 61  &     4.9  & 205$\pm$32 & 40$\pm$8  \\
 93  &     3.2  & 366$\pm$65 & 88$\pm$18 \\
\noalign{\smallskip}
\hline
\noalign{\smallskip}
\multicolumn{4}{c}{COBE-DIRBE} \\
\noalign{\smallskip}     
\hline
\noalign{\smallskip}
(THz)     & ($\mu$m)   & \multicolumn{2}{c}{Jy}\\
\noalign{\smallskip}     
\hline
\noalign{\smallskip}
  1.25 & 240    & 153000$\pm$3000 & 12350$\pm$1440 \\
  2.14 & 140    & 251100$\pm$4500 & 18900$\pm$2005 \\
  3.0  & 100    & 200200$\pm$3000 & 15750$\pm$2125 \\
  5.0  &  60    &  50500$\pm$7500 &  7010$\pm$730  \\
 12.0  &  25    &   7520$\pm$1100 &   425$\pm$70   \\
 25.0  &  12    &   3450$\pm$600  &   235$\pm$35   \\
 61.2  &   4.9  &   1240$\pm$190  &   150$\pm$25   \\
 85.7  &   3.5  &   2190$\pm$300  &   280$\pm$40   \\
136.4  &   2.2  &   3765$\pm$540  &   525$\pm$65   \\
240.0  &   1.25 &   4520$\pm$650  &   670$\pm$80   \\
\noalign{\smallskip}
\hline
\end{tabular}
\end{center}
\label{DIRBEdata}
\end{table}

The global spectral energy distribution (SED) of entire galaxies
provides an important tool to study global properties, such as star
formation activity, ISM heating and cooling balance, extinction and
dust content. Understanding the SEDs of nearby galaxies is essential
to the interpretation of measurements of very distant galaxies.
Especially important galaxies to study are the Large Magellanic Cloud
(LMC) and the Small Magellanic Cloud (SMC), the southern hemisphere
Milky Way satellites. They are so close (LMC: D = 50 kpc, SMC: D = 63
kpc) that we can study them in exhaustive detail allowing us to relate
global to local properties. However, their very proximity gives them
very large angular extent (LMC: 8$^{\circ}$, SMC: 2$^{\circ}$), so
that global flux densities have been determined only in the relatively
recent past, and over limited wavelength ranges. Particularly lacking
is coverage over a broad spectral range from the far infrared
(typically 2-3 THz or 0.10-0.14 mm) to the radio regime (typically
5-10 GHz or 3-6 cm).  This includes the mm/submm wave region that may
reveal unique information on properties and emission mechanisms of
dust as well the thermal free-free emission from ionised gas resulting
from the star formation processes.

Emission in this spectral range all but requires observatories in
space for its measurement.  The all-sky surveys by the cosmological
satellites WMAP and COBE provide a unique opportunity to acquire the
missing information and complement existing data.  In this paper, we
have used the COBE and WMAP archive databases to extract maps and
globally integrated flux densities for the Clouds. As we will describe
below, this resulted, among other things, in the surprising discovery
of a significant excess of emission from the Magellanic Clouds at
millimeter and sub-millimeter wavelengths.

\section{WMAP and COBE maps of the Magellanic Clouds}

\subsection{WMAP data}

The WMAP mission and its data products have been described in detail
by Bennett et al (2003a, b, c).  For our analysis of the Magellanic
Clouds, as in the case of Centaurus A (Israel et al. 2008), we used
the reduced and calibrated Stokes I maps of the entire sky from the
official WMAP 5-year release (Hinshaw et al. 2009).  The maps were
observed at frequencies $\nu$ = 22.5, 32.7, 40.6, 60.7, and 93.1 GHz
with resolutions of 53, 40, 31, 21, and 13 arc-min respectively.  The
HEALPIX data maps were converted to flat maps in Zenithal Equal Area
projection with pixel solid angles of
$p(o)\,=\,1.90644\,\times\,10^{-6}$ sr and intensities in mK.  We
integrated over the full extent of the LMC and the SMC and determined
the flux densities given in Table\,\ref{DIRBEdata} from the summed
values with conversion factors Jy/mK = 30.7 $\times$ $p(o)$ $\times$
$\nu^{2}$. In the case of the SMC, we took care to exclude the patch
of Milky Way foreground emission at position
-2$^{\circ}$,+1.5$^{\circ}$ in Fig.\,\ref{WMAPs}. In order to find
large-scale regional differences, we also determined flux densities
for sub-regions of the Magellanic Clouds, including the 30 Doradus
region in the LMC, and the SMC Wing; they are listed in
Table\,\ref{WMAPpeakdata}. The errors quoted are much larger than the
formal errors in the integrated flux densities, because we took into
account the uncertainty in the required integration area and the
uncertainty in the Milky Way foreground contribution.  The maps of
both Clouds are shown in Figs.\,1 and 2 to illustrate noise and
confusion levels.

\subsection{COBE data}

The COBE satellite was launched in 1989 to measure the diffuse
infrared and microwave radiation from the early universe (Boggess et
al., 1992). It carried three instruments, a Diffuse Infrared
Background Experiment (DIRBE), a Differential Microwave Radiometer
(DMR), and a Far Infrared Absolute Spectrophotometer (FIRAS).

\subsubsection{DIRBE}

The DIRBE experiment mapped the sky at wavelengths of 1.25 ($J$), 2.20
($K$) 3.5 ($L$) 4.9 ($M$), 12, 25, 60, 100, 140, and 240 $\mu$m with a
$0.7^{\circ}$ beam (Silverberg et al. 1993; see also Wall et
al. 1996). We extracted map data in Galactic coordinates for the parts
of the sky containing the LMC and the SMC, respectively, and reversed
both longitude and latitude directions in order to obtain images
corresponding to the sky distribution.  On the LMC and the SMC, we
extracted cubes of $51\times51$ and $41\times41$ pixels in extent,
respectively.  Each pixel is 0.3 degrees in size. These fields were
chosen because they are centred on the lowest believable contour in
the 100$\mu$m maps.  In each map, we subtracted both unrelated
(stellar) point sources and a smoothed background.  Flux densities,
determined by integrating over the relevant map areas, are listed in
Tables\,\ref{DIRBEdata} and \ref{DIRBEpeakdata}.
Fig.\,\ref{DIRBEmaps} shows the distribution of infrared luminosity
over the LMC and SMC, as derived from integrating over all DIRBE
channels, and also serves a reference for the positions listed in
Table\,\ref{DIRBEpeakdata}.  Individual channel maps are not shown
here\footnote{Foreground-corrected channel images derived from the
  Zodi-Subtracted Mission Average Maps can be found on Karl Gordon's
  website
  http://dirty.as.arizona.edu/~kgordon/research/mc/mc.html}.
The quoted errors are dominated by calibration uncertainties and are
significantly larger than the formal random errors.

\subsubsection{FIRAS}

The FIRAS experiment was designed for precise measurements of the
cosmic microwave background spectrum and to observe the dust and line
emission from the Galaxy. It covered the wavelength range from 0.1 to
10 mm in two spectral channels and had approximately 5$\%$ spectral
resolution and a 7$^{\circ}$ field of view (Wright et al. 1991; Fixsen
et al. 1994, 1997). We extracted a continuum spectrum of the LMC, the
SMC being too weak. Because the LMC was only marginally resolved by
FIRAS, we averaged the spectrum over all pixels within 6 degrees of
galactic coordinates (280.47, -32.89). From this, we subtracted an OFF
spectrum that was the average of pixels in the annulus from 7 to 12
degrees from the center. We do not list the results for the individual
204 spectral points, but they are shown in Fig.\,\ref{Magspec}.

%Table 2  WMAP Local Data
\begin{table*}
\caption[]{WMAP peak flux densities in the Magellanic Clouds}
\begin{center}
\begin{tabular}{cccccccccccc}
\hline
\noalign{\smallskip}
         &\multicolumn{8}{c}{LMC}&\multicolumn{3}{c}{SMC}\\
Pos$^{a}$ &1+2+3&4&6&7&8&9&10&12&1&2&3\\
Name$^{b}$ & 30Dor$^{c}$& N206&N77-94&Bar& N11 & N44 & N48  & N57  &Wing&NE&SW\\
\noalign{\smallskip}     
\hline
\noalign{\smallskip}
$\nu$ & \multicolumn{11}{c}{Flux densities} \\
(GHz)     & \multicolumn{11}{c}{Jy}\\
\noalign{\smallskip}     
\hline
\noalign{\smallskip}
 23      & 55 (72)  &  3.3 &  6.4 &  5.2 &  6.6 &  4.1 &  3.2 &  4.8 &  2.5 &  7 &  7\\
 31      & 57 (72)  &  3.6 &  6.6 &  5.2 &  7.5 &  4.5 &  2.9 &  5.1 &  3   &  9 &  8\\
 41      & 57 (71)  &  4.2 &  6.9 &  5.5 &  7.9 &  4.9 &  2.8 &  5.3 &  4   & 10 & 10\\
 61      & 58 (74)  &  6.7 &  8.7 &  6.5 & 10.5 &  6.8 &  2.8 &  6.7 &  8   & 16 & 16\\
 93      &101 (198) & 13.1 & 19.1 & 10.5 & 19.3 & 13.3 &  4.0 & 11.5 & 19   & 34 & 35\\
\noalign{\smallskip}
\hline
\end{tabular}
\end{center}
Notes a. Position numbers correspond to DIRBE position numbers in
Table\,\ref{DIRBEpeakdata}, but area covered is not
identical. b. Object identifies the most prominent feature within the
aperture, usually an HII region complex. c. Numbers in the first
column are flux densities with local background level subtracted;
numbers in the second column, between parentheses, are flux densities
as measured by integrating over the aperture without background
subtraction.
\label{WMAPpeakdata}
\end{table*}

%Table 3  COBE Local Data
\begin{table*}
\caption[]{COBE-DIRBE peak flux densities in the Magellanic Clouds}
\begin{center}
\begin{tabular}{cccccccccccccccc}
\hline
\noalign{\smallskip}
          &\multicolumn{12}{c}{LMC}&\multicolumn{3}{c}{SMC}\\  
Pos.$^{a}$ &1&2&3&4 &5 &6 &7 &8 &9 &10 &11 &12&1 & 2& 3 \\ 
Name$^{b}$&30Dor&N159&Ridge&N206&---&N77-94&Bar&N11&N44&N48&N63&N57&Wing&NW&SW\\
Pixel& 24,32& 22,32& 19,33& 18,30& 17,20& 22,19& 23,26& 31,18& 27,27& 33,27& 33,30& 29,30& - & - &-\\
\noalign{\smallskip}     
\hline
\noalign{\smallskip}
$\lambda$& \multicolumn{15}{c}{Flux Densities} \\
($\mu$m) & \multicolumn{15}{c}{kJy}\\
\noalign{\smallskip}     
\hline
\noalign{\smallskip}
  1.25& 0.174& 0.228&0.121&0.113&0.031&0.090& 0.315&0.053& 0.085&0.047&0.064&0.059&0.102&0.177&0.315 \\
  2.2 & 0.154& 0.196&0.098&0.088&0.022&0.076& 0.251&0.048& 0.066&0.040&0.059&0.046&0.077&0.133&0.238 \\
  3.5 & 0.116& 0.126&0.056&0.049&0.013&0.049& 0.143&0.032& 0.043&0.025&0.035&0.030&0.042&0.071&0.128 \\
  4.9 & 0.084& 0.077&0.031&0.028&0.008&0.030& 0.081&0.019& 0.027&0.015&0.018&0.019&0.023&0.040&0.072 \\
 12   & 0.78 & 0.41 &0.106&0.068&0.026&0.128& 0.21 &0.095& 0.140&0.063&0.040&0.081&0.028&0.060&0.091 \\
 25   & 3.16 & 1.08 &0.114&0.095&0.020&0.25 & 0.35 &0.21 & 0.29 &0.099&0.068&0.186&0.066&0.138&0.180 \\
 60   &24.9  &12.9  &1.56 &1.74 &0.307&3.05 & 6.04 &2.63 & 4.36 &1.43 &0.99 &2.92 &1.21 &2.74 &3.94  \\
100   &35.6  &22.1  &5.26 &4.23 &1.24 &7.21 &11.5  &6.39 & 9.04 &3.98 &2.69 &6.33 &3.06 &6.06 &8.51  \\
140   &39.0  &26.6  &7.92 &5.48 &2.33 &9.03 &13.1  &7.35 &11.1  &5.07 &3.20 &7.17 &3.62 &6.98 &9.47  \\
240   &19.2  &13.8  &5.32 &3.19 &2.04 &6.06 & 6.79 &4.68 & 6.71 &3.36 &2.30 &4.42 &2.68 &4.1  &5.18  \\
\noalign{\smallskip}     
\hline
\noalign{\smallskip}
&\multicolumn{12}{c}{Luminosity (10$^{7}$\,L$_{\odot}$)}\\
  1-240&15.6 &11.4  &3.32 &2.74 &0.90 &3.37 & 7.29 &2.44 & 4.23 &1.77 &1.57 &2.92 &3.21&5.68 &8.99  \\
 12-240&12.9 & 8.4  &1.79 &1.35 &0.43 &2.14 & 3.78 &1.74 & 3.07 &1.14 &0.82 &2.05 &1.52&2.86 &4.07  \\
 12-100& 9.0 & 5.3  &9.02 &0.75 &0.18 &1.23 & 2.36 &1.02 & 1.83 &0.62 &0.46 &1.22 &8.71&1.69 &2.44  \\
\noalign{\smallskip}
\hline
\end{tabular}
\end{center}
Note. a. Position numbers correspond to WMAP position numbers in
Table\,\ref{WMAPpeakdata}, but area covered is not identical. b. The
objects refer to the most prominent features included in the
aperture. These are the giant HII region complex 30 Doradus (1); the
very bright HII region complex N159 south of 30 Dor (2); the
prominent ridge of radiating dust and molecular gas south of 30 Dor
and N159 (3); the bright but isolated HII region complex N206 south of
the LMC Bar (4); a very extended HII region complex, containing N77,
N79, N83, N87, N90, N93, N94 at the southwestern corner of the LMC
(6); the LMC Bar itself (7); the second brightest HII region complex
N11 at the northwestern corner of the LMC (8); the third brightest HII
region complex N44 north of the Bar (9); the HII region complexes N48,
N63, and N57, delineating Shapley Constellation III (10, 11, 12)
\label{DIRBEpeakdata}
\end{table*}

%Figure 1, 2: WMAPs
\begin{figure*}[]
\unitlength1cm
\begin{minipage}[t]{3.6cm}
\resizebox{3.67cm}{!}{\rotatebox{0}{\includegraphics*{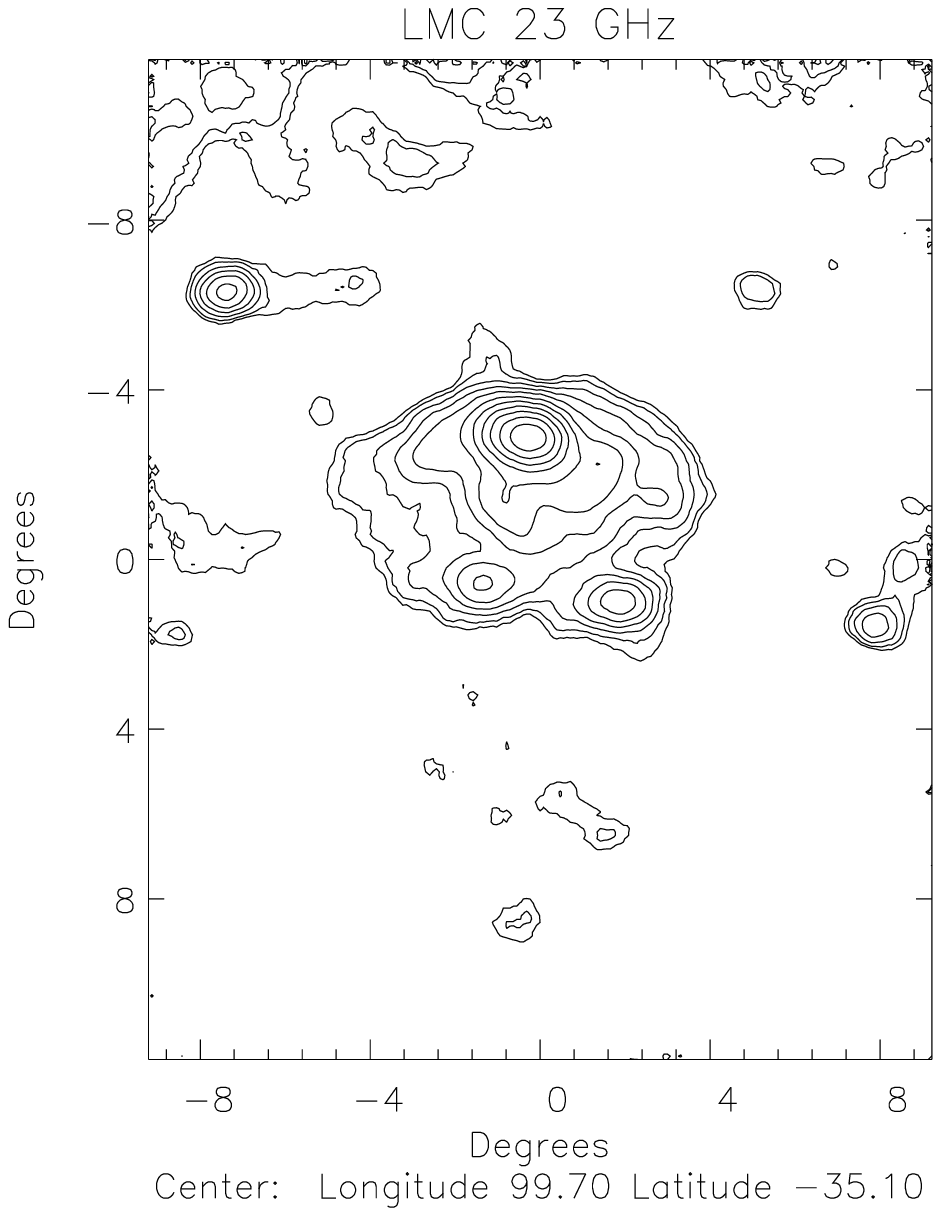}}}
\end{minipage}
\begin{minipage}[t]{3.6cm}
\resizebox{3.67cm}{!}{\rotatebox{0}{\includegraphics*{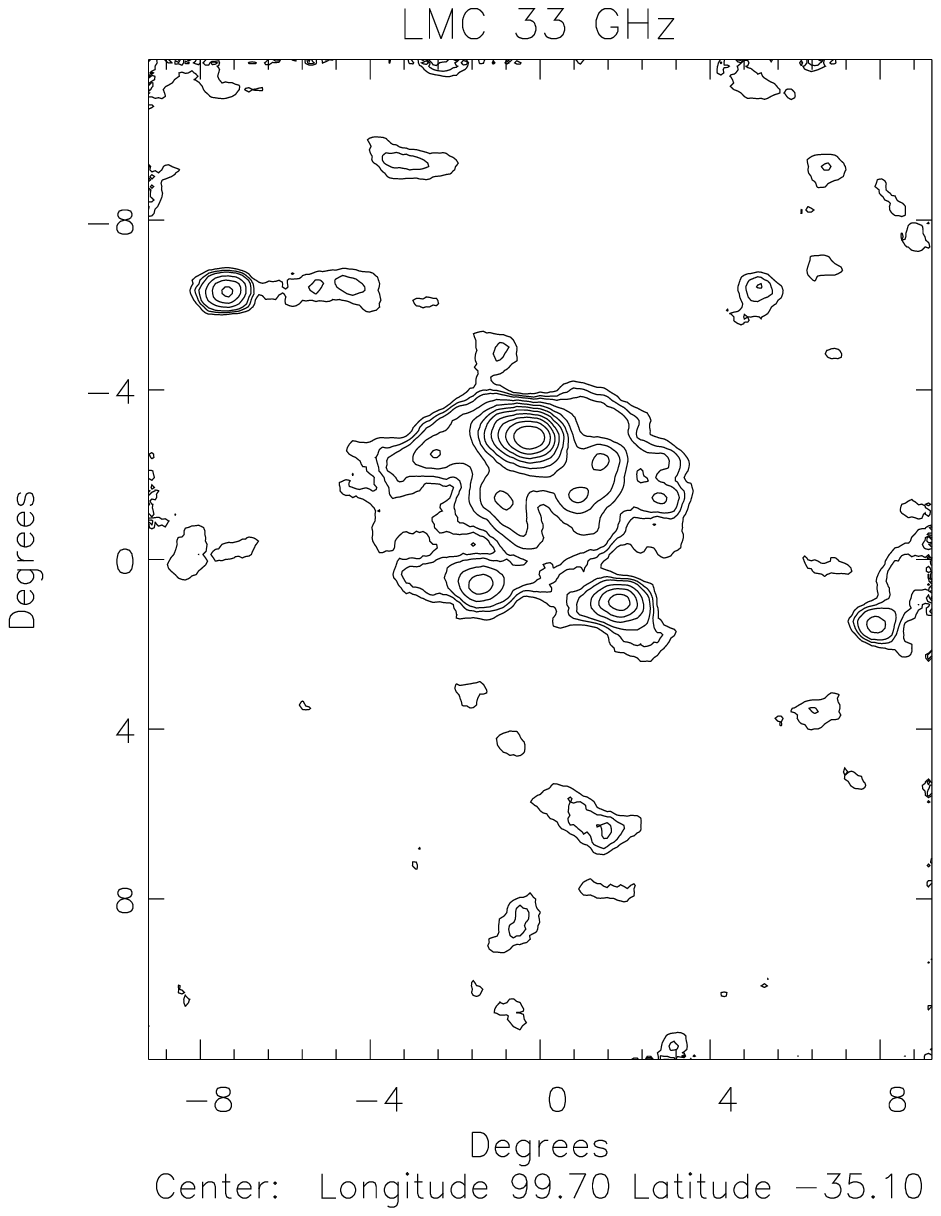}}}
\end{minipage}
\begin{minipage}[t]{3.6cm}
\resizebox{3.67cm}{!}{\rotatebox{0}{\includegraphics*{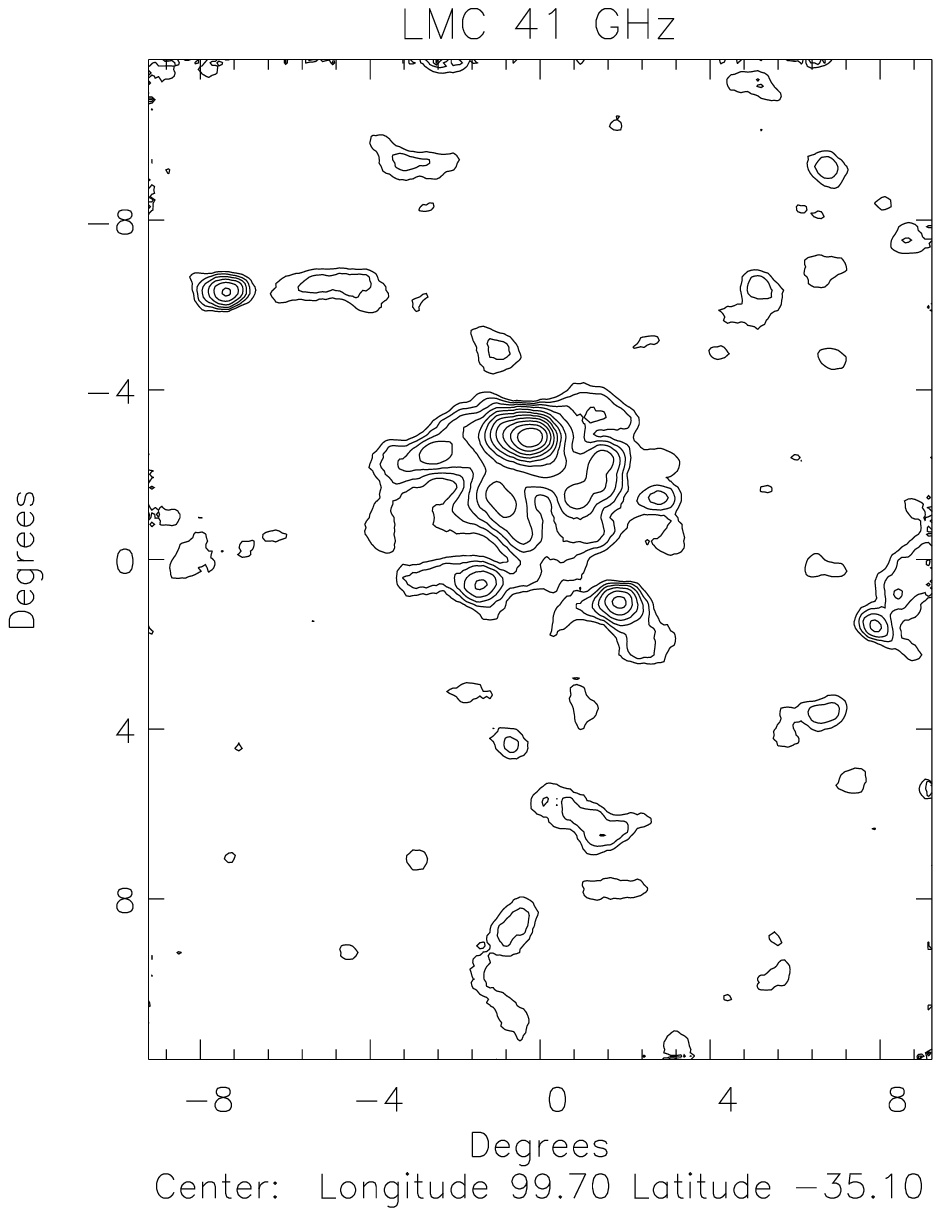}}}
\end{minipage}
\begin{minipage}[t]{3.6cm}
\resizebox{3.67cm}{!}{\rotatebox{0}{\includegraphics*{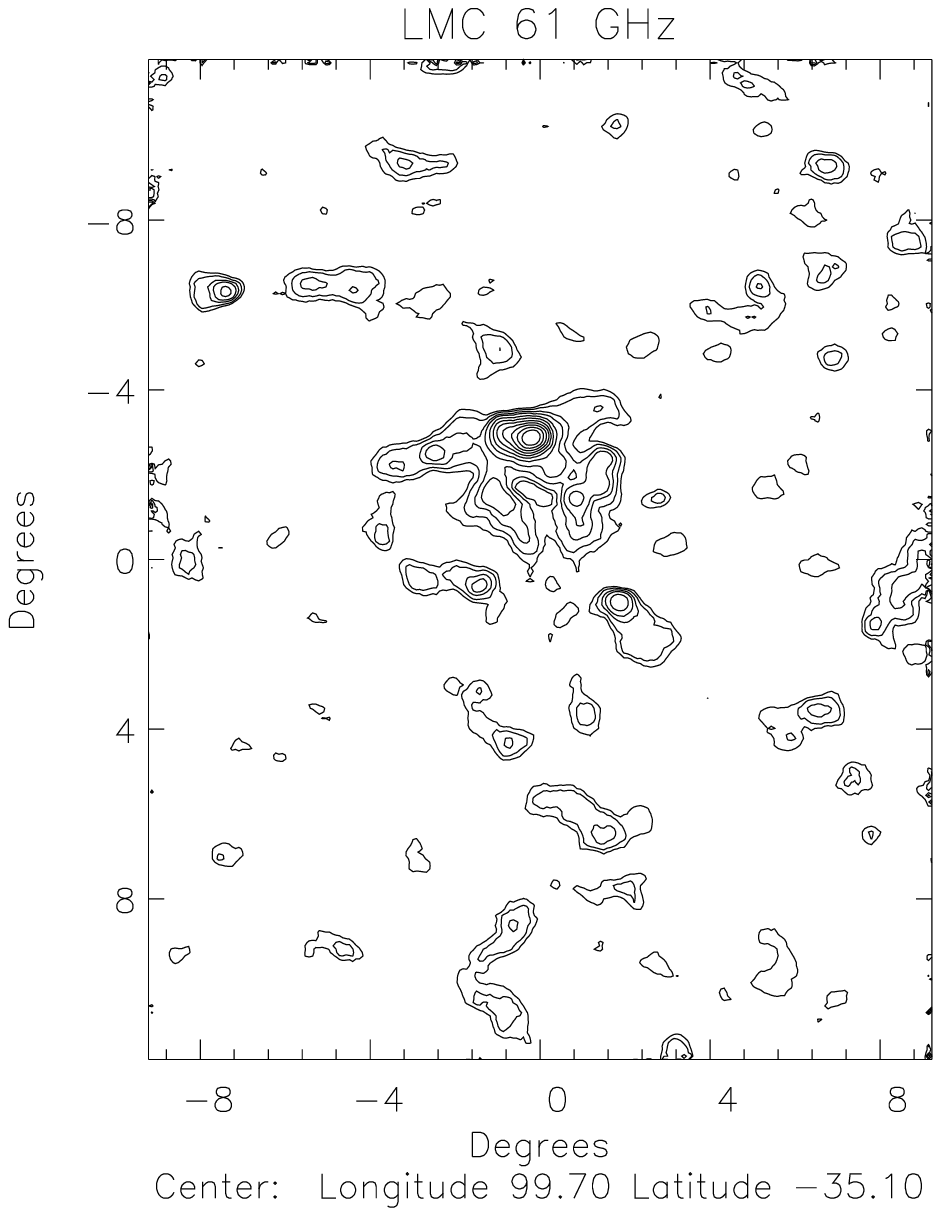}}}
\end{minipage}
\begin{minipage}[t]{3.6cm}
\resizebox{3.67cm}{!}{\rotatebox{0}{\includegraphics*{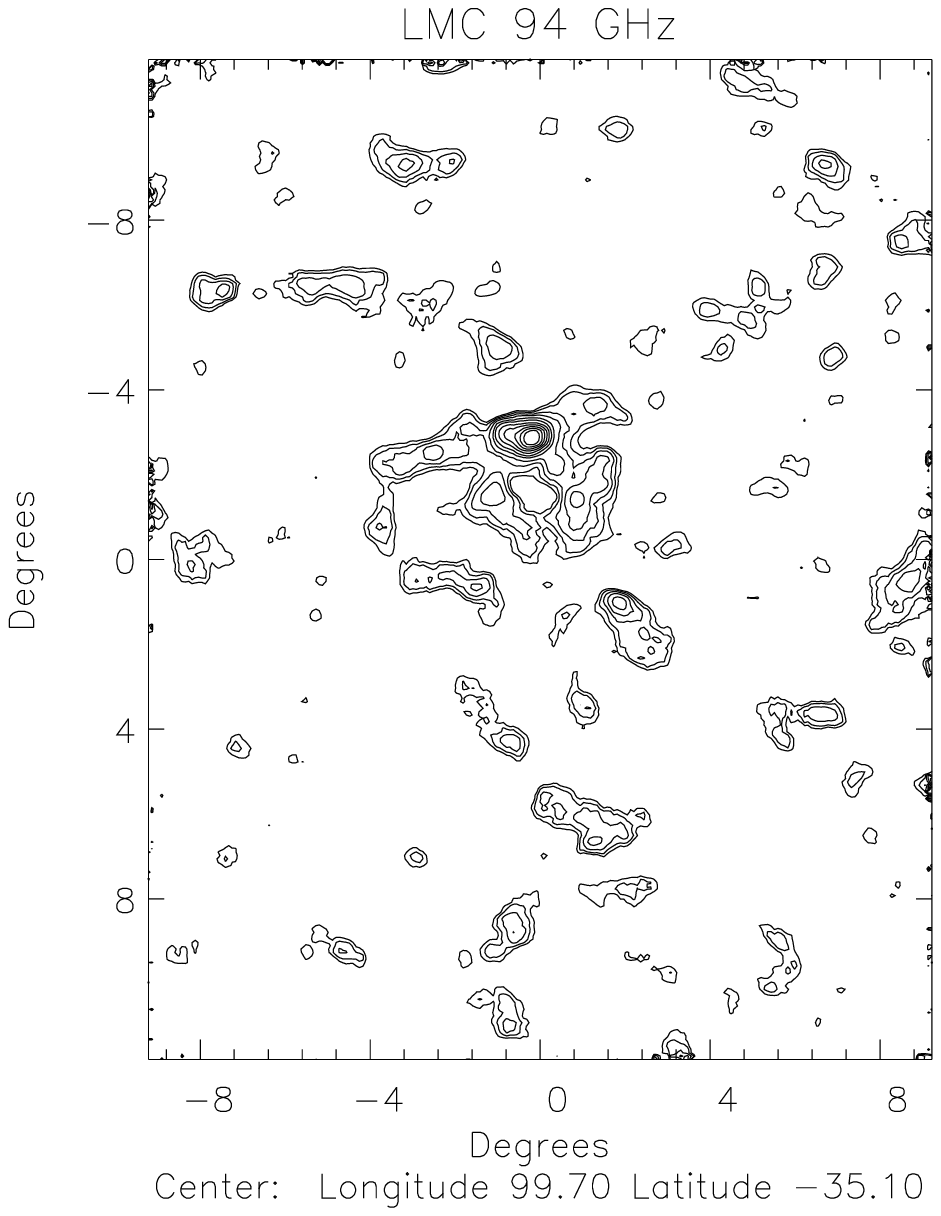}}}
\end{minipage}
\caption[]{Maps of the radio continuum emission of the LMC at (left to
  right) 23, 33, 41 GHz, 61 GHz, and 94 GHz.  All images are at the
  nominal WMAP resolution, and in Galactic coordinates centered on
  $l=279.70$, $b=-35.10$.  Equatorial North is at right. Contour
  levels are drawn at (23 GHz) 0.1, 0.19, 0.38, 0.74, 1.4, 2.8, 5.5,
  10.7 mK, (33 GHz) 0.1, 0.19, 0.36, 0.68, 1.3, 2.4, 4.6 mK; (41 GHz)
  0.1, 0.19, 0.34, 0.64, 1.2, 2.2, 4.1, 7.6 mK; (61 GHz) 0.1, 0.17,
  0.31, 0.53, 0.94, 1.6, 2.9, 5.0 mK; (93 GHz) 0.1, 0.16, 0.25, 0.40
  0.64 1.0, 1.6, 2.6 mK. }
\begin{minipage}[t]{3.6cm}
\resizebox{3.67cm}{!}{\rotatebox{0}{\includegraphics*{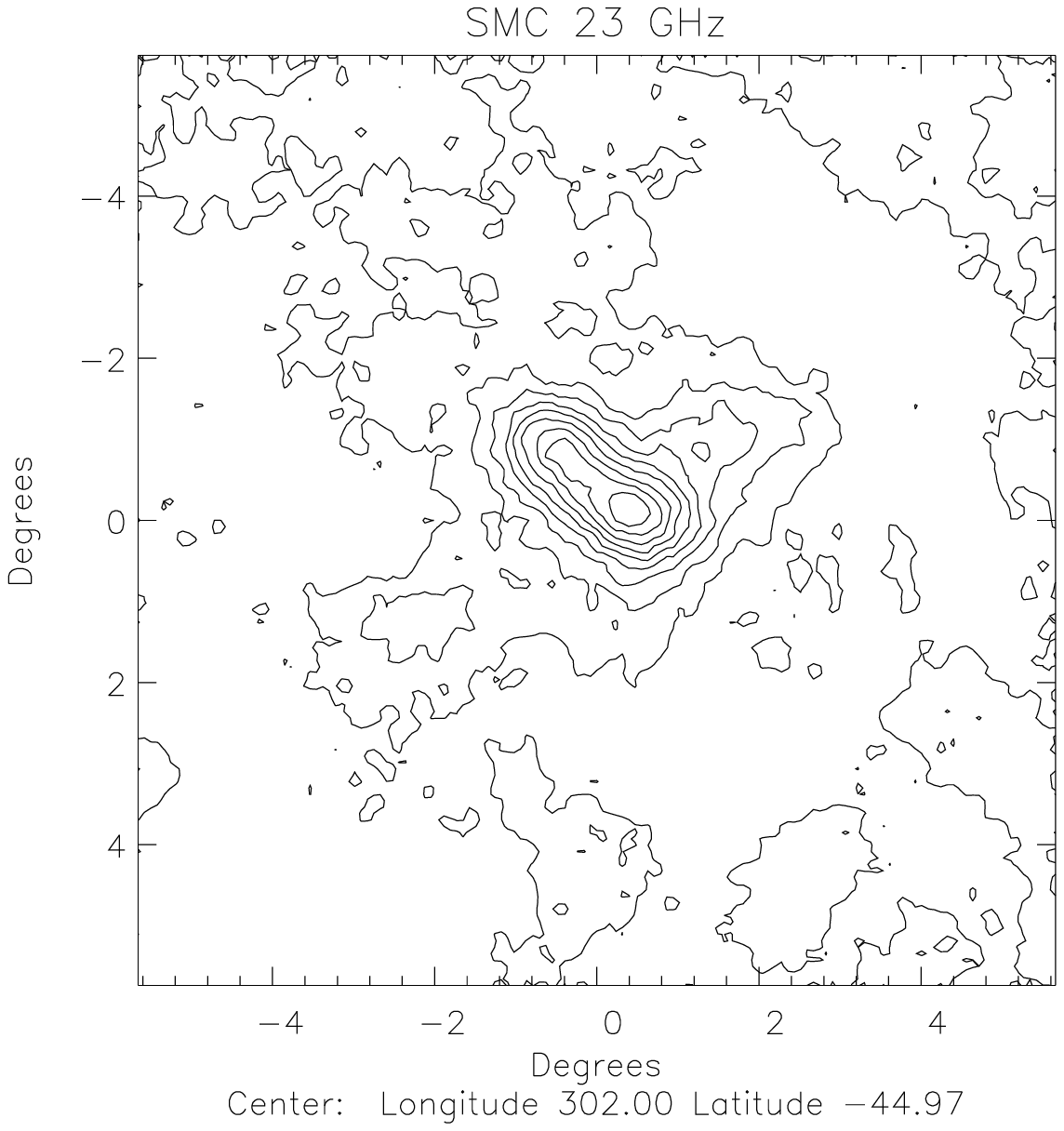}}}
\end{minipage}
\begin{minipage}[t]{3.6cm}
\resizebox{3.67cm}{!}{\rotatebox{0}{\includegraphics*{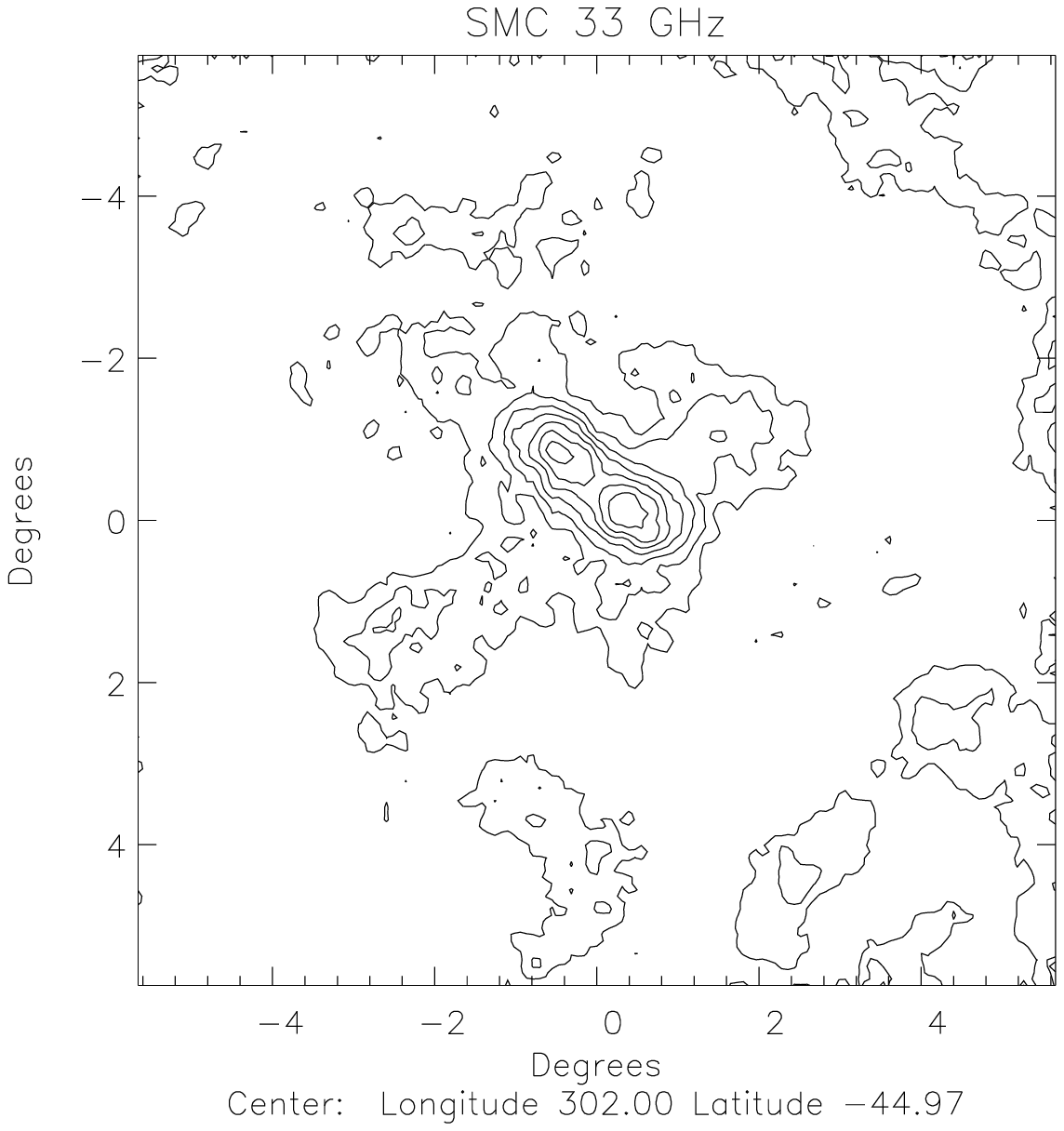}}}
\end{minipage}
\begin{minipage}[t]{3.6cm}
\resizebox{3.67cm}{!}{\rotatebox{0}{\includegraphics*{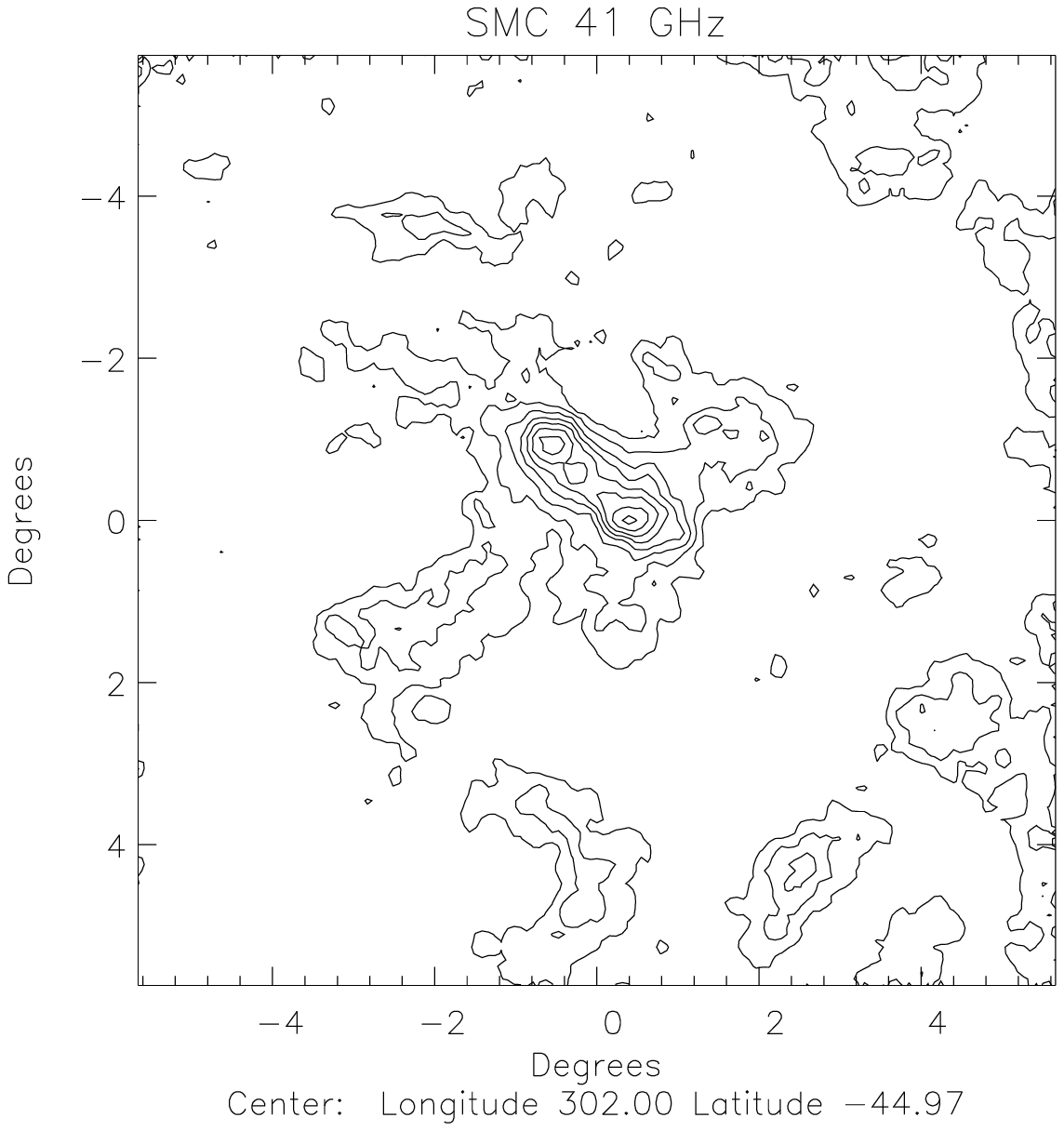}}}
\end{minipage}
\begin{minipage}[t]{3.6cm}
\resizebox{3.67cm}{!}{\rotatebox{0}{\includegraphics*{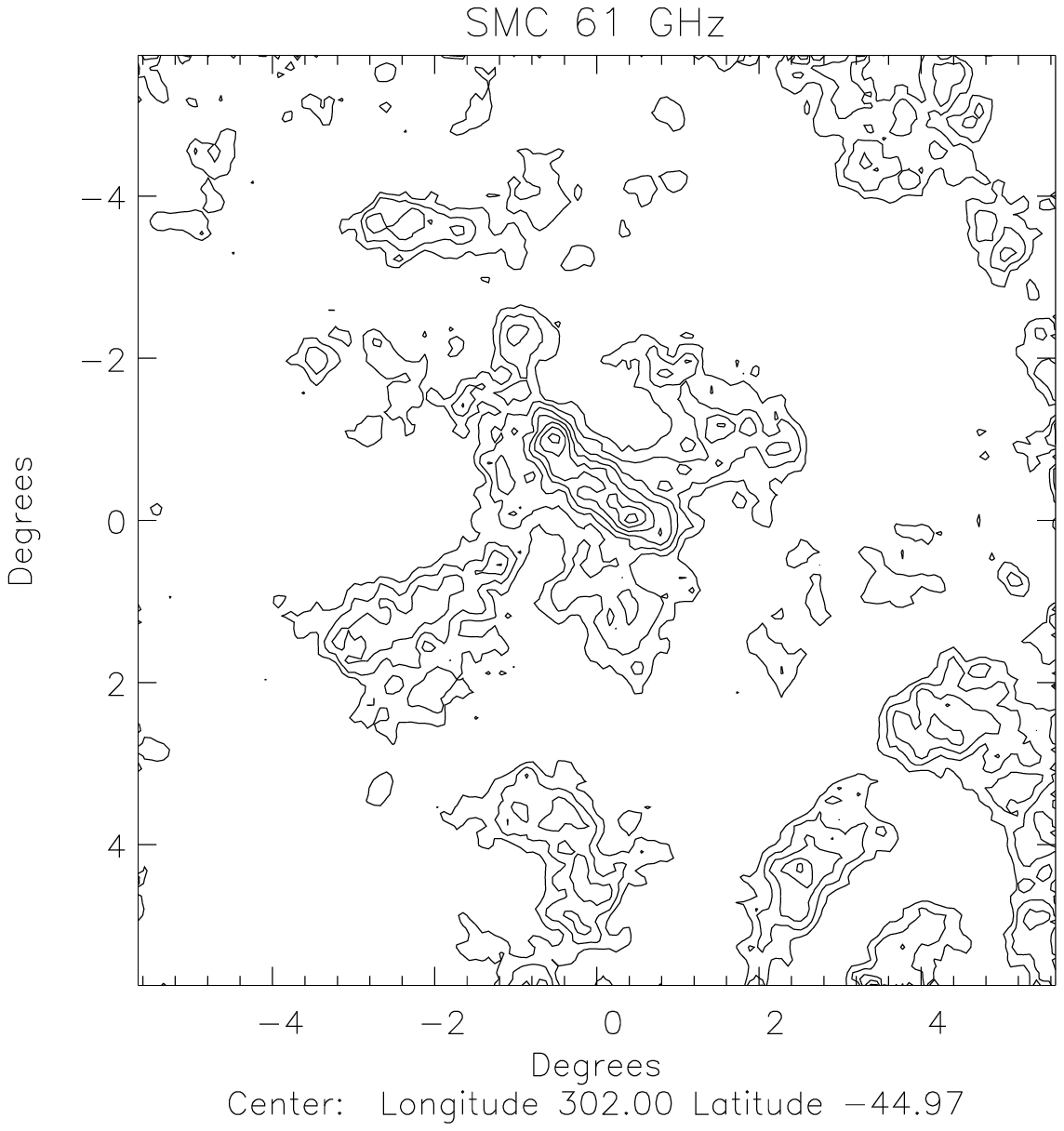}}}
\end{minipage}
\begin{minipage}[t]{3.6cm}
\resizebox{3.67cm}{!}{\rotatebox{0}{\includegraphics*{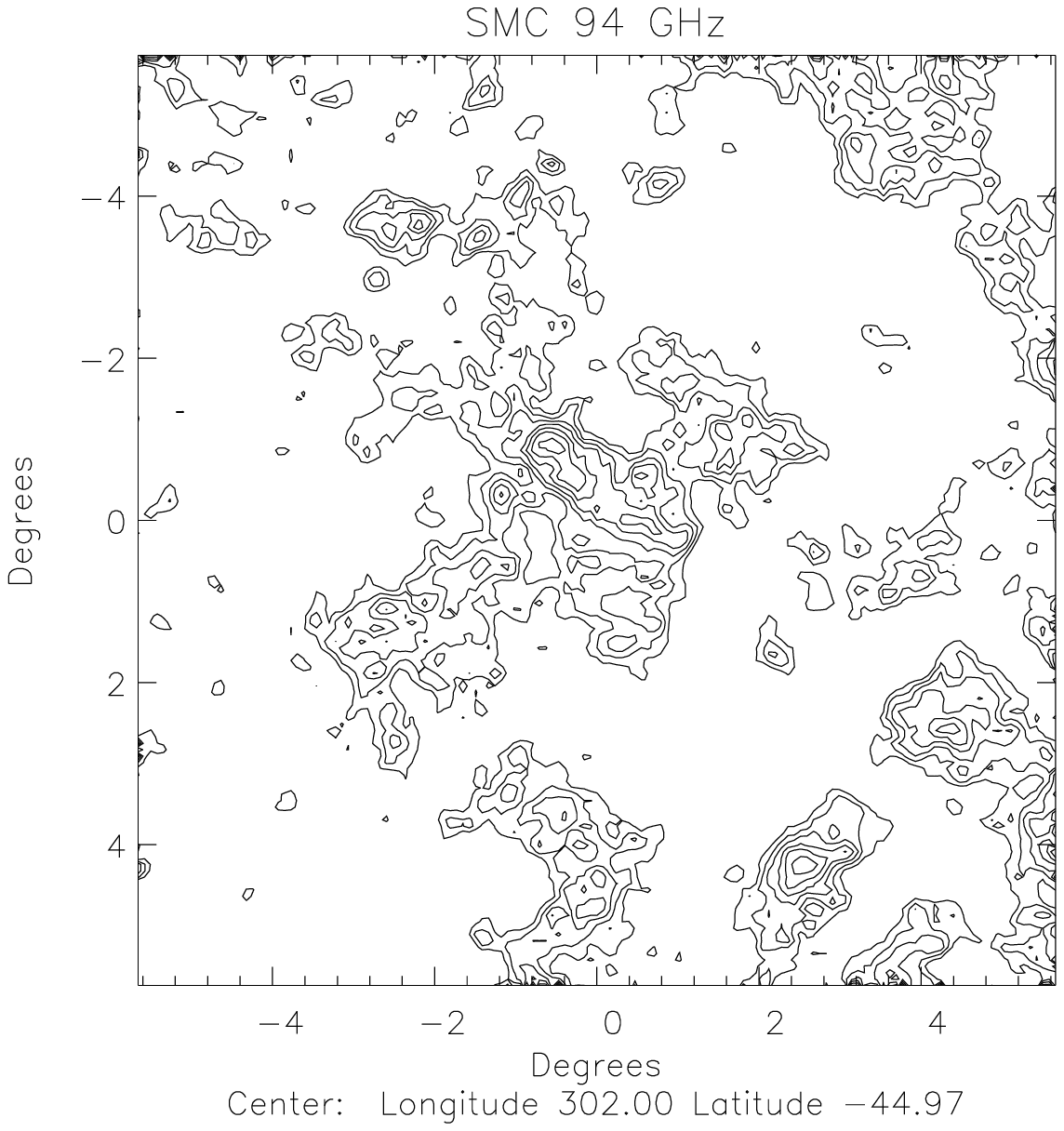}}}
\end{minipage}
\caption[]{Maps of the radio continuum emission of the SMC at (left to
  right) 23, 33, 41 GHz, 61 GHz, and 94 GHz.  All images are at the
  nominal WMAP resolution. Images are in Galactic coordinates,
  centered on $l=302.00$, $b=-44.97$, and therefore appear
  `upside-down'; equatorial North is at bottom.  Contour levels are
  drawn at (23 GHz) 0.1, 0.19, 0.38, 0.74, 1.4, 2.8, 5.5, 10.7 mK, (33
  GHz) 0.1, 0.19, 0.36, 0.68, 1.3, 2.4, 4.6 mK; (41 GHz) 0.1, 0.19,
  0.34, 0.64, 1.2, 2.2, 4.1, 7.6 mK; (61 GHz) 0.1, 0.17, 0.31, 0.53,
  0.94, 1.6, 2.9, 5.0 mK; (93 GHz) 0.1, 0.16, 0.25, 0.40 0.64 1.0,
  1.6, 2.6 mK. }
\label{WMAPs}
\end{figure*}

%Figure 3 DIRBE Maps
\begin{figure*}[]
\begin{minipage}[t]{18.0cm}
\resizebox{9.251cm}{!}{\rotatebox{90}{\includegraphics*{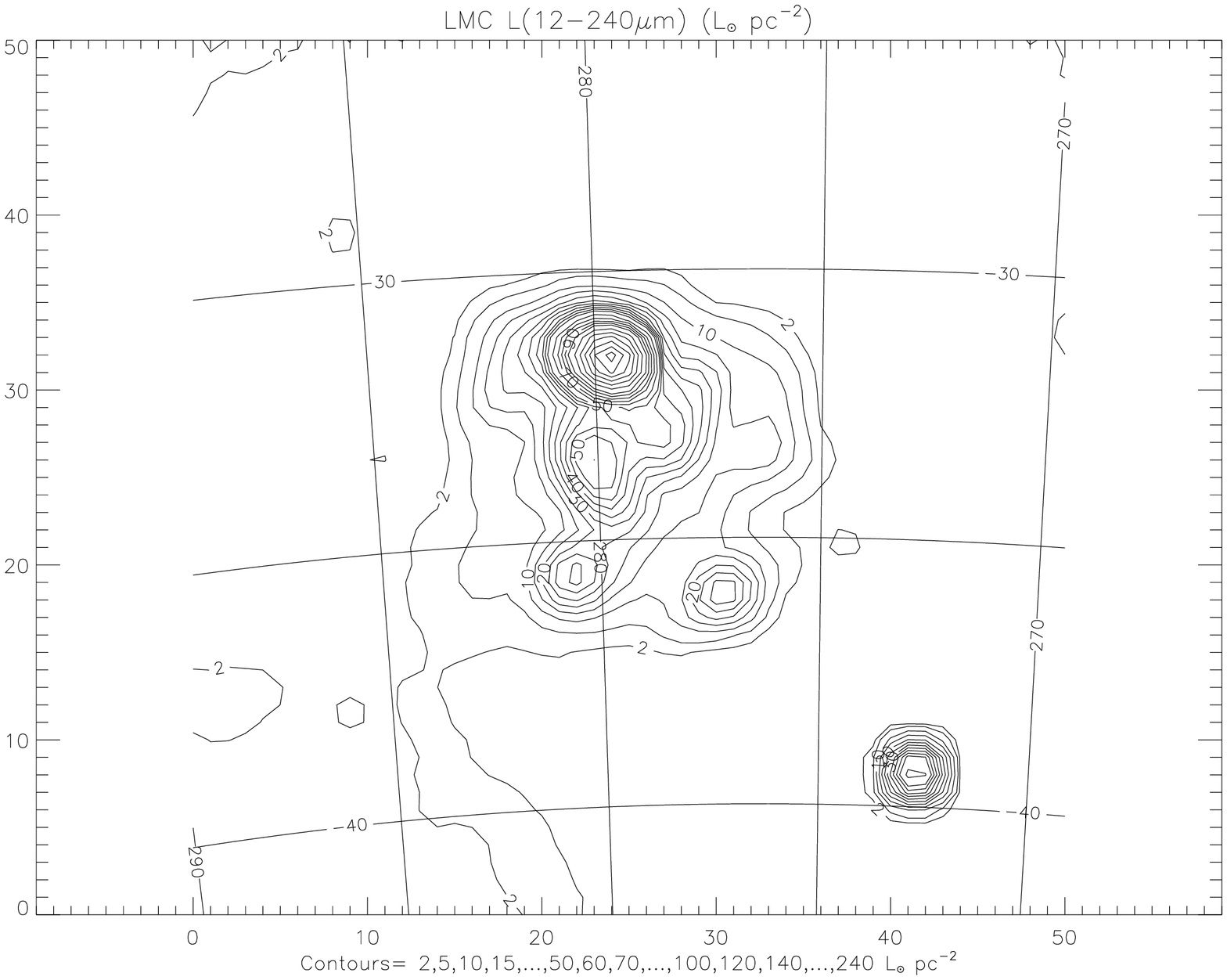}}}
\resizebox{9.251cm}{!}{\rotatebox{90}{\includegraphics*{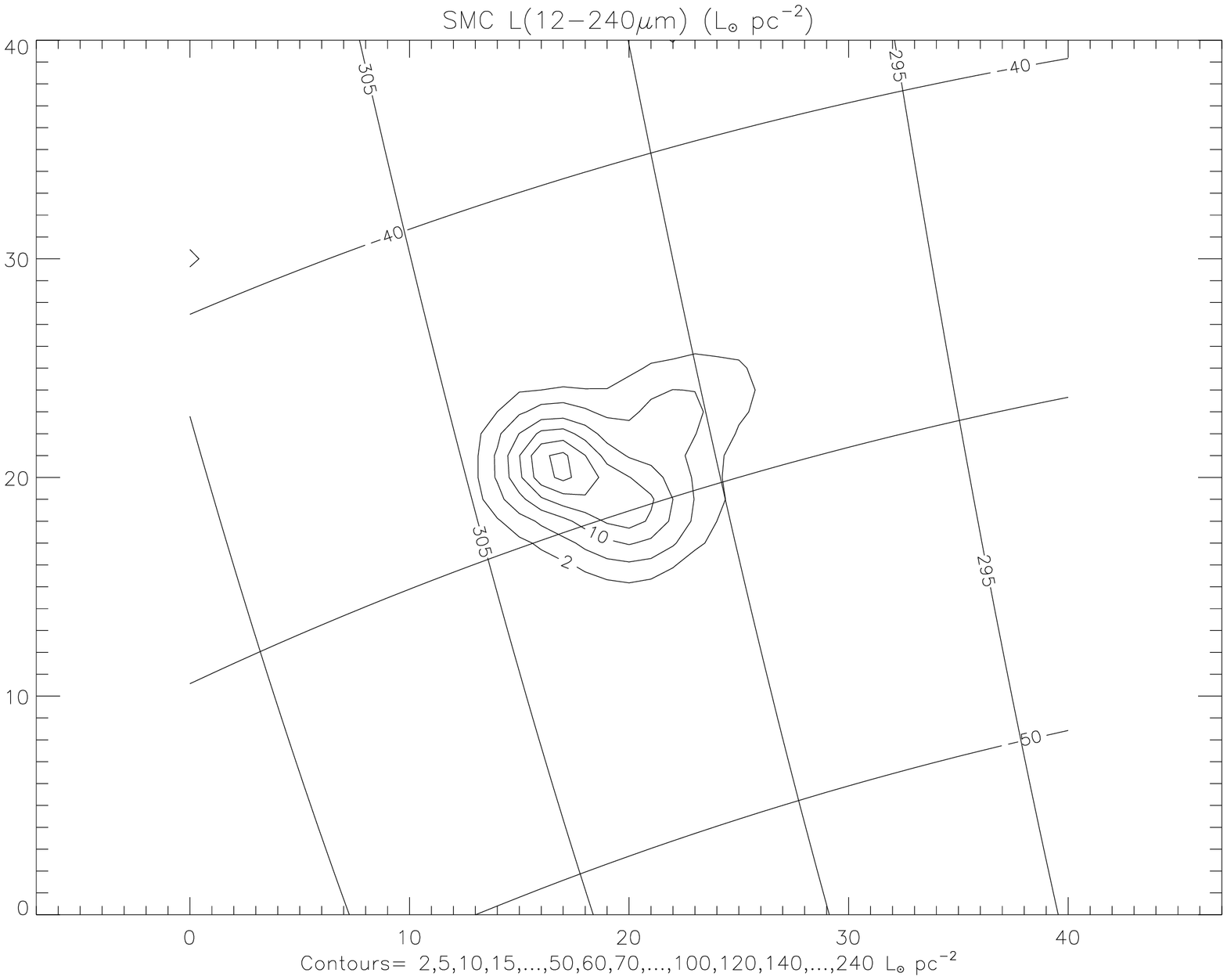}}}
\end{minipage}
\caption[] {Total infrared luminosity maps from DIRBE
  measurements. Contours are marked in units of
  L$_{\odot}$pc$^{-2}$. Galactic coordinates are indicated. Left:
  LMC. The bright star-forming regions 30 Doradus, N11, and the
  complex of HII regions N77-N94 are easily identified as is the Bar
  (cf. Table 3).  Bright object at pixel (41,8) is the Galactic
  foreground star R Dor.  Right: SMC. The pixel positions in
  Table\,\ref{DIRBEpeakdata} refer to the $x$ and $y$ axis values}
\label{DIRBEmaps}
\end{figure*}

\subsection{Literature data} 
\label{maglitdat} 

We have used the IPAC NED as a guide to find LMC and SMC global flux
densities over a wide frequency range in the published literature.
The NED compilation should be used with care, as it is not complete and
also includes values of limited spatial coverage often significantly
underestimating total flux densities.  Consequently, we also searched
the recent literature, and in all cases took care to select only those
flux densities that (a) were reliably determined, and (b) correspond
to the entire galaxy, not just the bright circumnuclear region.

In addition to the integrated optical flux densities $U_{T}$, $B_{T}$,
and $V_{T}$ (i.e. not corrected for extinction) from the Third
Reference Catalog of Bright Galaxies, (RC~3 -- de Vaucouleurs et
al. 1991), we have used radio data obtained or collected by Loiseau et
al. (1987), Mountfort et al. (1987), Alvarez et al. (1987, 1989),
Klein et al (1989), Ye $\&$ Turtle (1991), and Haynes et al. (1991).
Infrared and sub-millimeter continuum data were taken from Schwering
1988, Rice et al. (1988), Stanimirovic et al. (2000), Aguirre et
al. (2003), Wilke et al. (2004), Hughes et al. (2006), Bolatto et
al. (2007), and Leroy et al. (2007).  The ultraviolet data of the LMC
are those of Page and Carruthers (1981). All data used are listed in
the on-line appendix.

Comparing the Magellanic Cloud measurements from the WMAP and the
DIRBE surveys (Table\,\ref{DIRBEdata}) with published results from
other spacecraft (IRAS, ISO, Spitzer) surveys, we note that the
infrared peak intensities measured by the COBE-DIRBE experiment for
the SMC are very close to these, but somewhat higher for the LMC.  As
the large extent of the LMC renders its measurement with relatively
small beams more sensitive to base-level uncertainties, we prefer the
low-resolution DIRBE flux-densities.  The radio continuum spectrum is
uncertain below 200 MHz, but well-defined at the higher
frequencies. The high-frequency radio continuum measurements agree
well with the results from WMAP.

\subsection{Comparison galaxies}

In the following, we will compare the Magellanic Cloud results to
those of other galaxies. However, there are very few galaxies whose
emission in the sub-millimeter to centimeter wavelength range has been
sampled with a sufficient degree of accuracy and completeness. A more
or less exhaustive sample consists of the starburst (SB) disk galaxies
NGC~253, M~82, and NGC~4945 (also measured by WMAP, see Chen $\&$
Wright 2009), the (ultra)luminous infrared galaxies (ULIRG) Arp~220,
Mk~231, NGC~3690, and NGC~6240, and the star-forming blue compact
dwarf (BCDG) galaxies He~2-10, 2Zw40, NGC~4194, and NGC~5253.  The
SEDs of these three groups of galaxies will be discussed in more
detail in a forthcoming paper.

%Figure 4: LMC/SMC Total Spectrum and SED
\begin{figure*}[]
\begin{minipage}[t]{18.0cm}
\resizebox{9cm}{!}{\rotatebox{270}{\includegraphics*{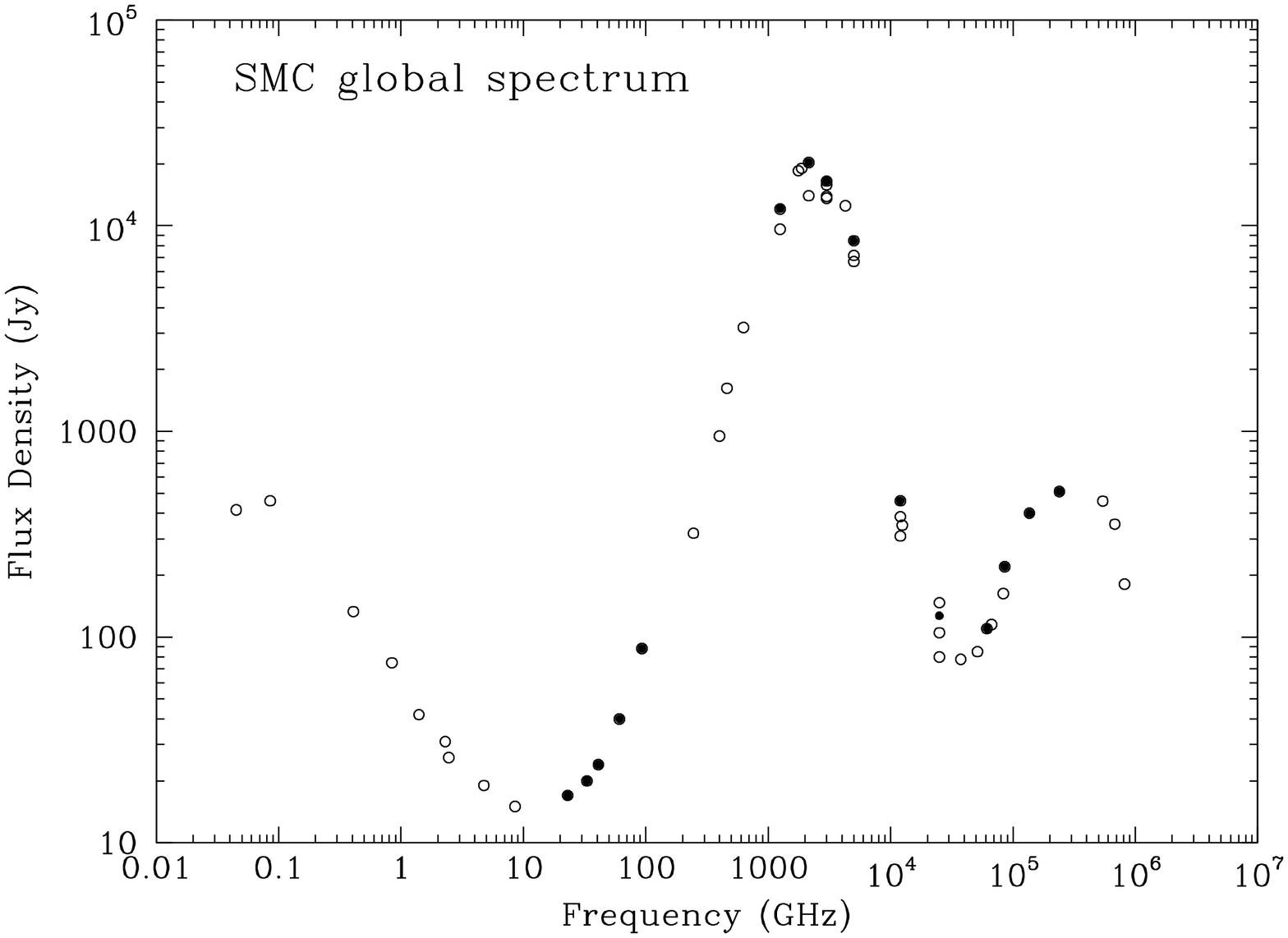}}}
\resizebox{9cm}{!}{\rotatebox{270}{\includegraphics*{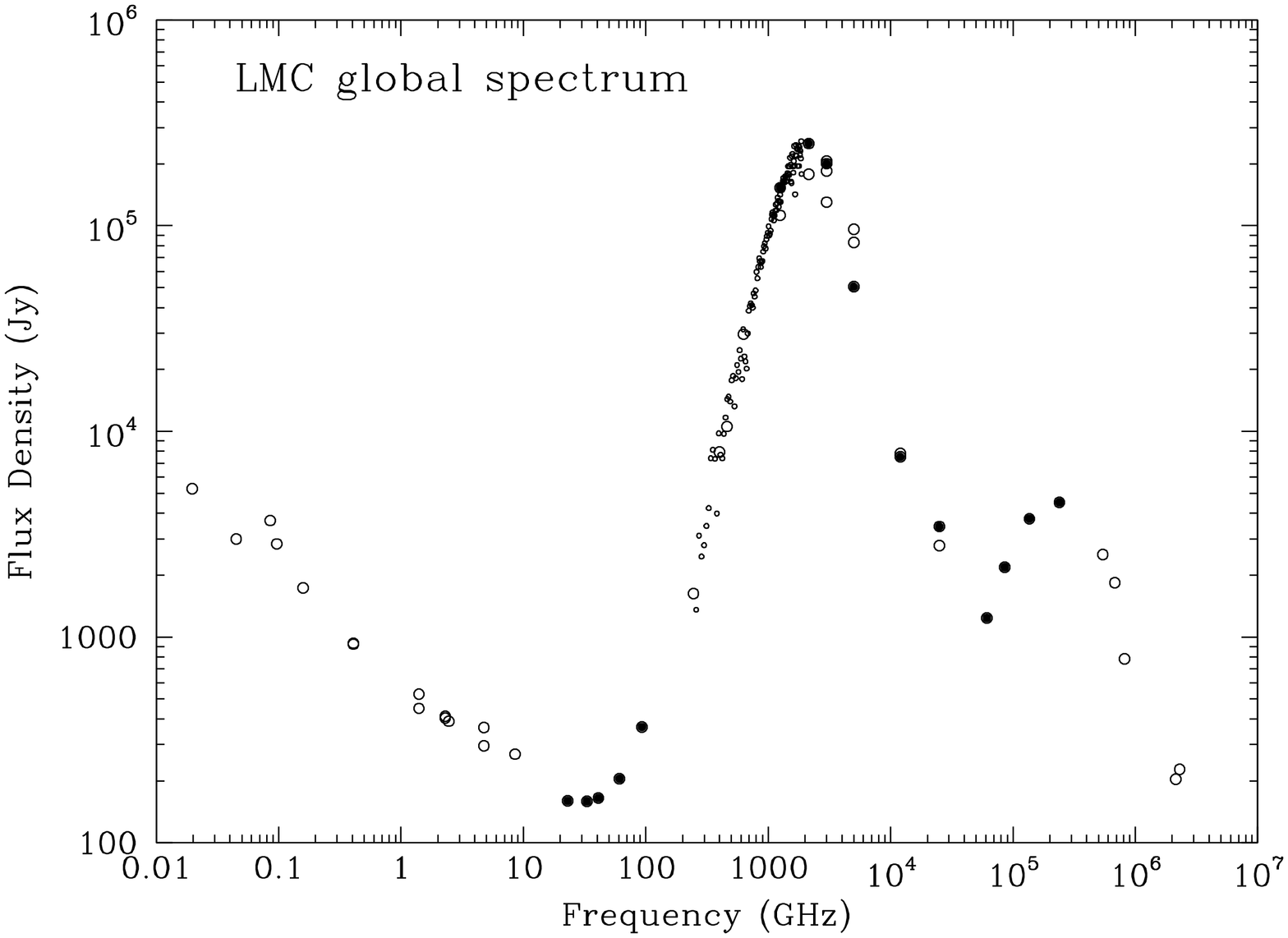}}}
\end{minipage}
\begin{minipage}[t]{18.0cm}
\resizebox{9cm}{!}{\rotatebox{270}{\includegraphics*{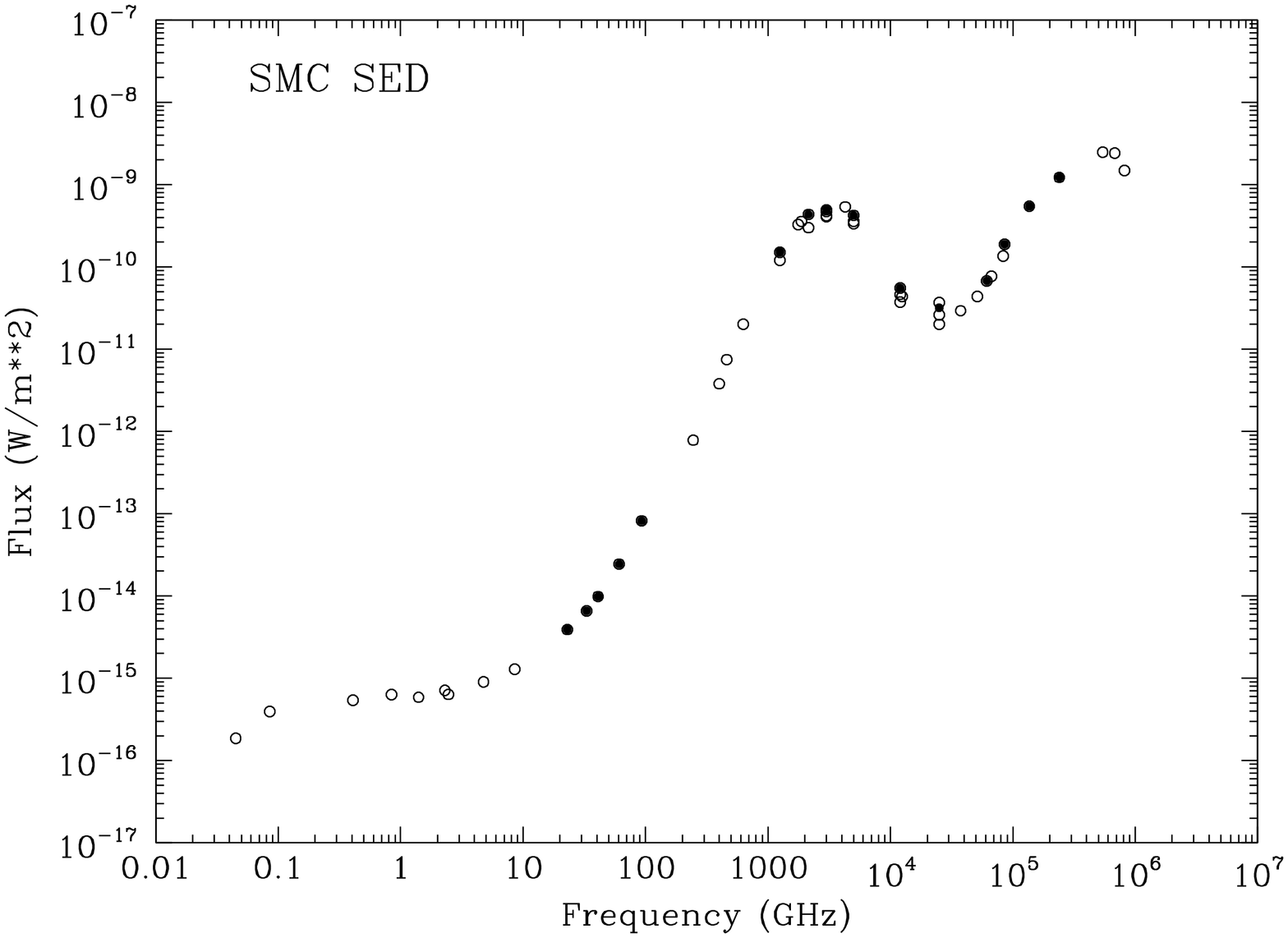}}}
\resizebox{9cm}{!}{\rotatebox{270}{\includegraphics*{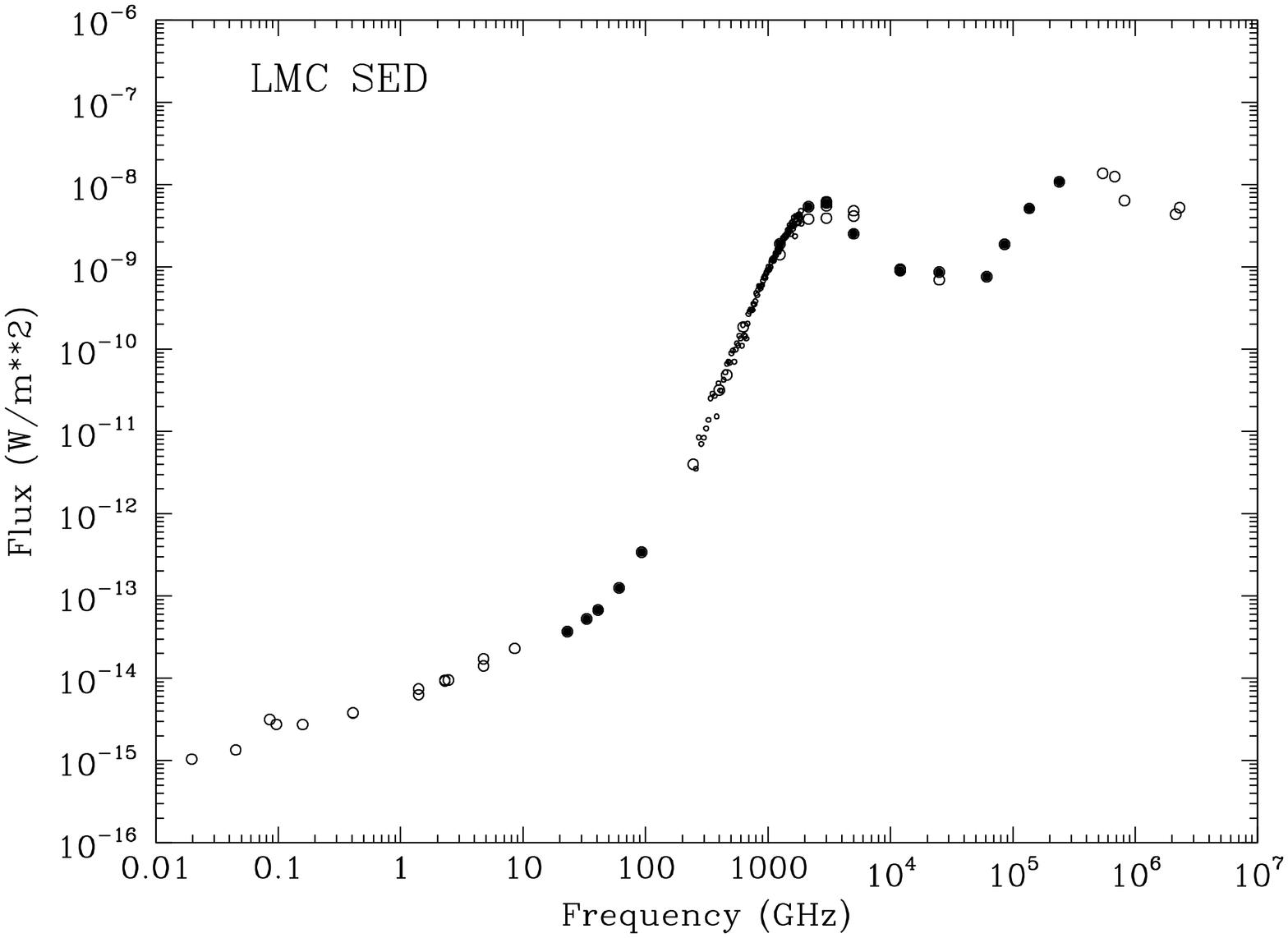}}}
\end{minipage}
\caption[] {Top: Global (area-integrated) continuum spectrum from
  low-frequency radio to the ultraviolet. Filled circles represent
  integrated flux densities in Jansky from Table\,\ref{DIRBEdata},
  open circles were taken from the literature (see text). Bottom:
  Corresponding global flux (proportional to power) distribution $\nu
  F_\nu$ in W m$^{-2}$. Direct starlight is stronger than starlight
  reemitted by dust, which is the reverse of the usual case in late
  type galaxies}
\label{Magspec}
\end{figure*}

\section{Results and analysis}

\subsection{Mid-infrared excess emission}

%Figure 5: DIRBE spectra
\begin{figure}[]
\begin{minipage}[t]{9.0cm}
\resizebox{9.2cm}{!}{\rotatebox{0}{\includegraphics*{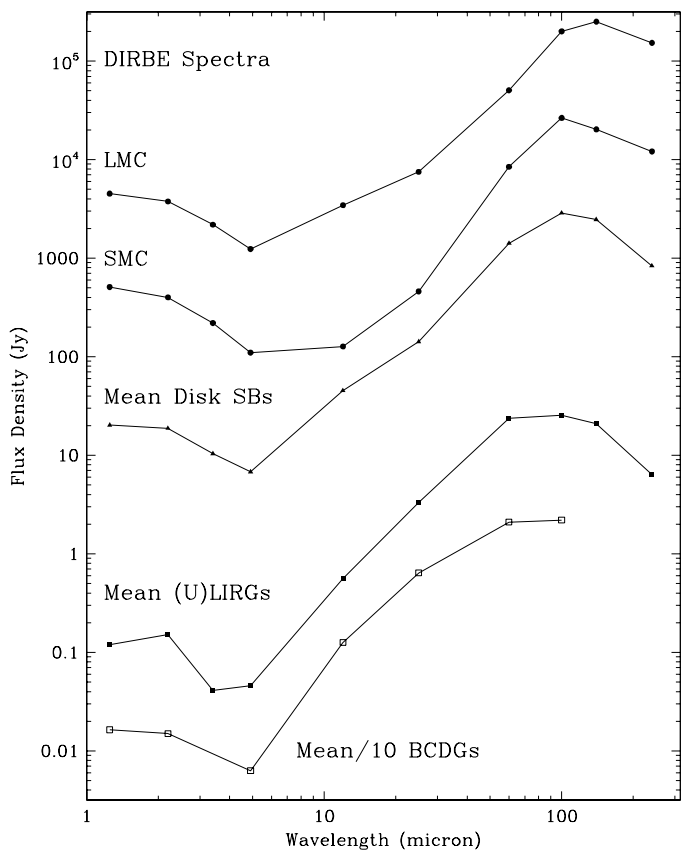}}}
\end{minipage}
\caption[] {Near- to far-infrared spectra of the LMC and the SMC,
  compared to a sample of star-burst galaxies taken from the DIRBE
  point source catalog by Smith et al. 2004 and data tabulated by Hunt
  et al. 2005. Most of the galaxies exhibit a clear 12$\mu$m excess,
  which is clearly lacking in the SMC and only weakly present in the
  LMC.}
\label{galdirbe}
\end{figure}

%Figure 6: Local Specs
\begin{figure}[]
\begin{minipage}[t]{9.0cm}
\resizebox{4.43cm}{!}{\rotatebox{0}{\includegraphics*{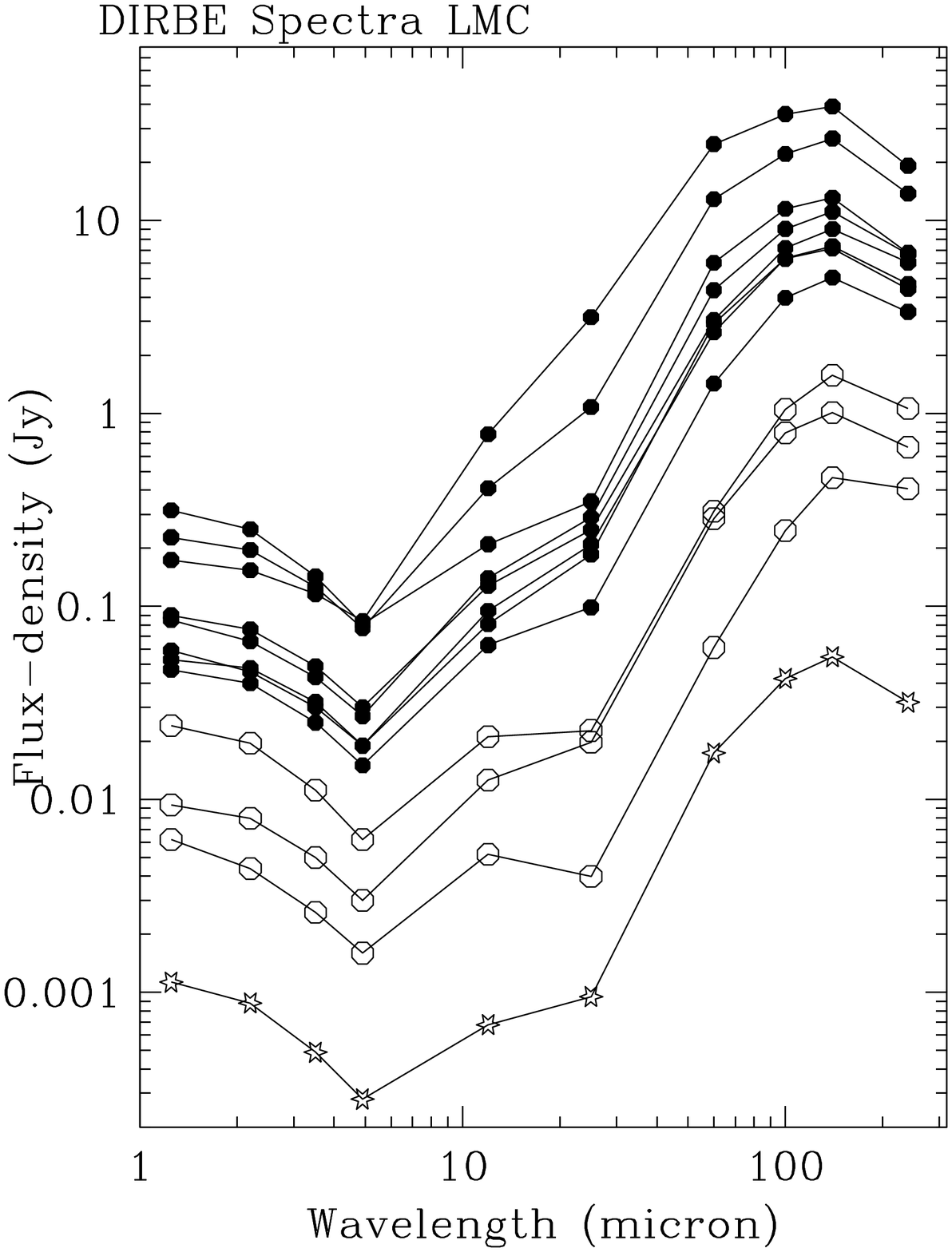}}}
\resizebox{4.43cm}{!}{\rotatebox{0}{\includegraphics*{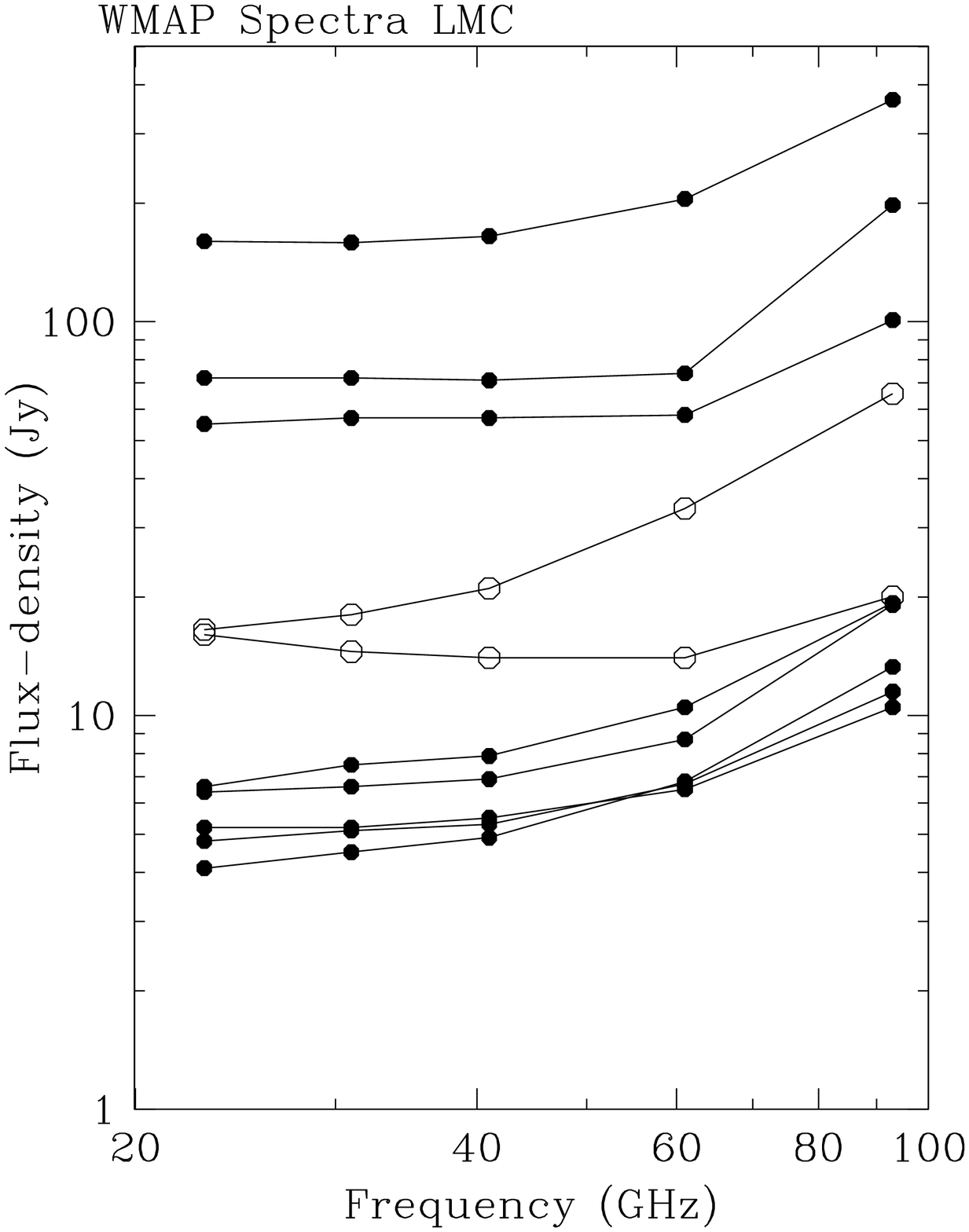}}}
\end{minipage}
\caption[] {COBE-DIRBE spectra of the LMC sub-regions from
  Table\,\ref{DIRBEpeakdata}. For the sake of clarity, we have divided
  the spectra (open circles) of LMC regions 3 (Dor Ridge), 5, and 10
  (N~48) by five, and that of LMC region 4 (N~206; open stars) by a
  hundred.  Right: WMAP spectra of the LMC sub-regions from
  Table\,\ref{WMAPpeakdata}).  The spectra of the northern source N~48
  and the southern object N~206, representing the most extreme cases,
  have been multiplied by five and are indicated by open circles. }
\label{localspec}
\end{figure}

From the observed flux density distributions we have determined the
mid-infrared excess emission in the 12$\mu$m broadband by logarithmic
interpolation between the observed 4.9$\mu$ and 25$\mu$ flux densities
($S_{12}$/$S_{12'}$, with log $S_{12'}$ = 0.56 log $S_{25}$ + 0.44
log$S_{5}$). Although more precise spectroscopy and photometry has
been provided by e.g. the Spitzer Space Observatory (see for instance
Draine et al. 2007; Bernard et al. 2008), we include here a brief
discussion of the DIRBE data because they allow us to compare the
integrated and large scale properties of the Magellanic Clouds to
those of other galaxies.  Fig.\ref{galdirbe} shows that there is no
mid-infrared excess in the SMC probably due to decreasing PAH
strengths (see e.g. Bolatto et al. 2007). Compared to other galaxies
there is only a weak excess in the LMC.  The sub-regions of the LMC
shown in Fig.\,\ref{localspec} reveal the 12$\mu$m excess to cover a
modest but non-negligible range.  Low metallicities, hard radiation
fields, and strong shocks may destroy the (8$\mu$m emitting ionised)
PAHs in all but the best-shielded locations (Engelbracht et al. 2005;
Micelotta 2009).  However, the 12$\mu$m excess may also be related to
the presence or absence of very small grains -- see, for instance, the
mid-infrared emission studies of the LMC by Sakon et al. (2006) and by
Bernard et al. (2008).  Fig.\,\ref{MIREDIRBE} shows that the
mid-infrared excess is {\it anti-correlated} with the energy density
of the radiation field as represented by the 100$\mu$m/140$\mu$m flux
density ratio (consistent with conclusions by Beir\~ao et al. (2006)
from NGC~5253 Spitzer data).  The coolest LMC regions in
Fig.\,\ref{MIREDIRBE} (numbers 3 and 5 in Table\,\ref{DIRBEpeakdata})
have the highest 12$\mu$m excess. As the low-metallicity SMC points
and the solar-metallicity galaxy points straddle the intermediate
metallicity LMC points, it appears that the mid-infrared excess is
proportional to metallicity in addition to being inversely
proportional to the radiation field.

\subsection{Radio spectrum and extinction}
\label{thermalextinction}

\subsubsection{Thermal radio contribution}

The LMC and SMC spectra in Fig.\,\ref{Magspec} and
Fig.\,\ref{localspec} both show a smooth transition from the radio to
the infrared with a broad minimum at 20--40 GHz (8--15 mm wavelength)
occurring as thermal emission from ionised gas becomes important
before the thermal emission from heated dust starts to dominate the
energy distribution.
 
The LMC and SMC radio spectra have spectral indices in the 0.1--5.0
GHz frequency range $\alpha_{LMC}\,\approx\,-0.55$ and
$\alpha_{SMC}\,\approx\,-0.63$ respectively (Alvarez et al. 1987,
1989; Loiseau et al. 1987; Klein et al. 1989; Haynes et al. 1991) with
$S_{\nu}\,\propto\,\nu^{\alpha}$.  Such values suggest significant
thermal contributions at the highest observed frequencies, but these
have been difficult to determine accurately because observations were
limited to the spectral range dominated by the non-thermal component.
The integrated H$\alpha$ fluxes measured by Kennicutt et al. (1995)
together with average {\it foreground} extinctions
$A_{V}(LMC)\,=\,0.25$ mag and $A_{V}(SMC) \,=\,0.12$ mag (Schlegel et
al, 1998) place {\it lower limits} on the thermal contributions of
$S_{10GHz}(LMC)\,\geq\,100$ Jy and $S_{10GHz}(SMC)\,\geq\,10$ Jy.  The
spectra presented in Fig.\,\ref{Magspec} show that these lower limits
are close to the observed total (thermal and non-thermal) flux densities
$S_{10GHz}(LMC)\,\approx\,175$ Jy and $S_{10GHz}\,\approx\,14.5$ Jy.

The present data allow us to separate with very high accuracy the
thermal and non-thermal contributions by fitting simultaneously the
radio spectrum over {\it both} the range where non-thermal emission
dominates {\it and} the range where thermal emission is dominant. The
LMC is best fit by a thermal continuum (spectral index
$\alpha\,=\,-0.1$) corresponding to $S_{10GHz}(th,LMC)\,=\,145\pm15$
Jy. The corresponding 5 GHz thermal fraction is 0.53. For the SMC we
find a best fit $S_{10GHz}(th,SMC)\,=\,13.4\pm1.0$ Jy, with
corresponding 5 GHz thermal fraction of 0.71.  Thus, we find in the
LMC a substantially lower and in the SMC a substantially higher
thermal contribution than estimated by Haynes et al. (1991) from
fitting the non-thermally dominated decimeter/centimeter radio data
only. Our fits also provide values for the spectral index of the
non-thermal emission, $\alpha_{LMC}\,=\,-0.70\pm0.05$ and
$\alpha_{SMC}\,=\,-1.09\pm0.10$.  The Lyman-continuum fluxes
log$N_{L}$ = 52.57 (LMC) and log$N_{L}$ = 51.72 (SMC) of the ionising
star ensembles in the Clouds implied by these thermal flux-densities
correspond to the presence of at least 2200 and 300 early O (mean
spectral type O6.5, cf. Vacca et al. 1996) in the LMC and the SMC
respectively.

\subsubsection{Extinction}

The free-free radio emission just determined and the integrated
H$\alpha$ fluxes ($f_{H\alpha}(LMC)\,=\,9.0\,\times\,10^{-8}$ erg
cm$^{-2}$ s$^{-1}$ and $f_{H\alpha}(SMC)\,=\,1.0\,\times\,10^{-8}$ erg
cm$^{-2}$ s$^{-1}$) from Kennicutt et al.  (1995) accurately
define the extinction of the Magellanic Clouds at the wavelength of
H$\alpha$ ($R$-band).  Including foreground, we find
$A_{H\alpha}(LMC)$ = $0.65\pm0.15$ mag and $A_{H\alpha}(SMC)$ =
$0.45\pm0.05$ mag.  This is in excellent agreement with the extinction
determined for samples of individual HII regions by Bell et al. (2002:
mean $A_{H\alpha}(LMC)\,=\,0.62\pm0.05$ for 52 objects) and Caplan et
al. (1996: mean $A_{H\alpha}(SMC)$ = $0.50\pm0.05$ for 35 HII
objects).  It follows that the extinction of the diffuse H$\alpha$
emission in both the LMC ($35\%$) and the SMC ($41\%$) (Kennicutt et
al. 1995) should also be very similar to the overall extinction.  When
we correct for the Milky Way foreground, the mean internal $R$-band
extinctions are $0.45\pm0.15$ mag (LMC) and $0.36\pm0.12$ (SMC) mag.
The very low mean extinctions of the Clouds are also obvious from the
-- for late-type galaxies -- unusually low ratios of the far-infrared
and optical peaks in their SEDs, ${{\nu S_{\nu}^{dust}}\over{\nu
    S_{V}^{star}}}\,\approx\,0.2$ (Fig.\,\ref{Magspec}).

%Figure 7: MIREDIRBE
\begin{figure}[]
\begin{minipage}[t]{9.0cm}
\resizebox{9.20cm}{!}{\rotatebox{270}{\includegraphics*{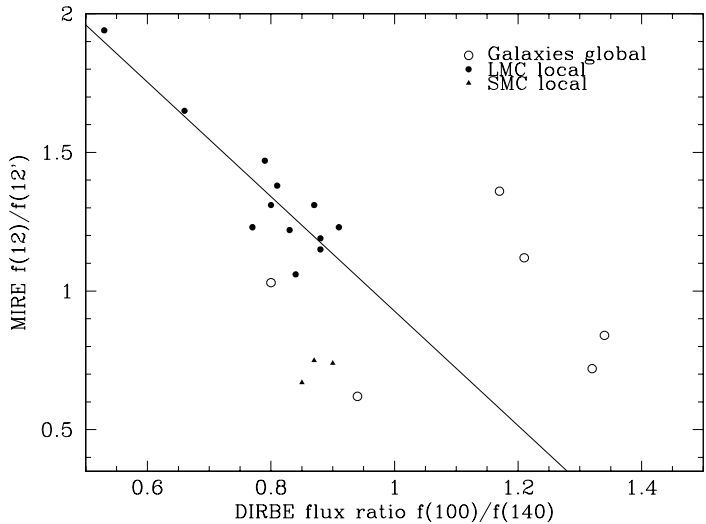}}}
\end{minipage}
\caption[] {The mid-infrared excess defined from the COBE-DIRBE
  measurements (see text) as a function of radiation field energy
  density. Open symbols represent the global emission from the LMC,
  the SMC and galaxies taken from the DIRBE point source catalog
  (Smith et al. 2004).  Filled symbols represent sub-regions in the
  LMC and the SMC, taken from Table\,\ref{DIRBEpeakdata}). The
  straight line is the linear regression for the LMC sub-regions.}
\label{MIREDIRBE}
\end{figure}

%Table 4  Millimeter excess
\begin{table}
\caption[]{Millimeter Excess}
\begin{center}
\begin{tabular}{lccccc}
\hline
\noalign{\smallskip}
Freq. & TD  & FF  & Syn & Excess & Total\\
(GHz) & \multicolumn{5}{c}{Jy}\\
\noalign{\smallskip}     
\hline
\noalign{\smallskip}     
\multicolumn{6}{c}{LMC: $T_{d}$ = 25.0 K; $\beta$ = 1.33}\\
245 & 1519 &  99 & 12  &   0 & 1630 \\
150 &  328 & 104 & 13  &  55 &  500 \\ 
93  &   64 & 109 & 18  & 175 &  366 \\
61  &   17 & 113 & 24  &  51 &  205 \\
33  &    2 & 121 & 36  &   0 &  159 \\ 
\noalign{\smallskip}     
\hline
\noalign{\smallskip}     
\multicolumn{6}{c}{SMC: $T_{d}$ = 29.5 K; $\beta$ = 0.91}\\
245 &  311 & 9.0 & 0.3 &   0 &  320 \\
150 &   81 & 9.5 & 0.5 &  69 &  160 \\ 
93  &   19 & 10.0& 0.9 &  58 &   88 \\
61  &  6.1 & 10.4& 1.4 &  23 &   40 \\
33  &  0.8 & 11.1& 1.7 &   7 &   20 \\ 
\noalign{\smallskip}
\hline
\end{tabular}
\end{center}
\label{Magexcess}
Note: TD = Thermal dust emission extrapolated from the spectral fit
published by Aguirre et al. (2003); FF = free-free thermal emission;
Syn = synchrotron nonthermal emission; Excess = emission in excess
over the sum of the preceding contributions, required to make up the
observed total emission listed in the last column (the 150 GHz flux
being an interpolation). The values for $T_{d}$ and $\beta$ were taken
from Aguirre et al. (2003).
\end{table}

\subsection{Millimeter and sub-millimeter excess in the Magellanic Clouds}

A major difference between the Magellanic Clouds and the WMAP
star-burst galaxies is a significant excess of emission at millimeter
and sub-millimeter wavelengths. In the latter, dust emission is the
dominant flux contributor only above 100 GHz (short-wards of 3 mm).
In contrast, the spectral upturn associated with dust emission occurs
in the LMC and especially the SMC at {\it much lower} frequencies of
30 GHz (7.5 mm) and 10 GHz (3 cm) respectively (see
Fig\,\ref{galwmap}).

%Figure 8: Comparison WMAP fluxes
\begin{figure}[]
\begin{minipage}[t]{9.0cm}
\resizebox{9.cm}{!}{\rotatebox{0}{\includegraphics*{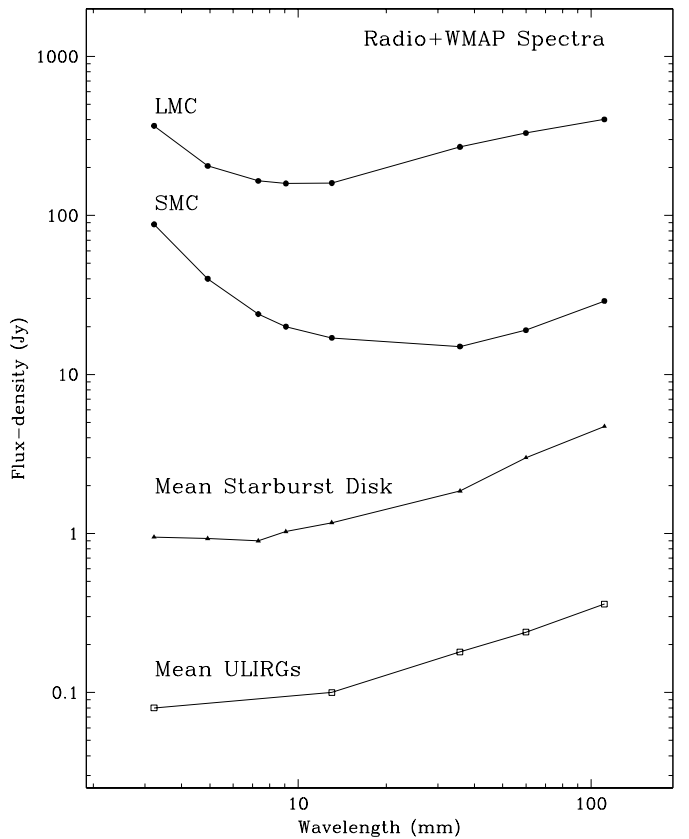}}}
\end{minipage}
\caption[] {Millimeter continuum spectra of the Magellanic Clouds
  compared to the means of other galaxies taken from the WMAP point
  source catalog (Chen $\&$ Wright 2009). WMAP points between 1.3 cm
  (23 GHz) and 3.2 mm (93 GHz) are supplemented by radio continuum
  data from 11 cm (2.7 GHz) to 3.6 cm (8.4 GHz) taken from the
  literature.  The spectra of the LMC and the SMC clearly exhibit an
  upturn extending to relatively long wavelengths (low frequencies)
  due to the presence of anomalous dust emission. In contrast, the
  spectra of the star-burst disk galaxies and those of the much more
  luminous (U)LIRGs show only spectral flattening as free-free
  emission becomes dominant at short wavelengths. }
\label{galwmap}
\end{figure}

\subsubsection{LMC and SMC millimeter and sub-millimeter spectra}

At sub-millimeter wavelengths, excess emission is known to occur in
dwarf galaxies such as NGC~1569 (Lisenfeld et al. 2002; Galliano et
al. 2003), II~Zw~40, He~2-10, NGC~1140 (Galliano et al. 2005), as well
as the star-burst galaxies NGC~3310 (Zhu et al. 2009), and NGC~4631
(Dumke et al. 2004; Bendo et al. 2006). In all these cases, there is a
relatively small excess of emission at wavelengths of 0.85 and 1.2 mm
over values extrapolated from the far-infrared peak assuming a big
grain emissivity $\beta$ = 2 (where $\beta$ is defined by
$F_{\nu}\,\propto\,B_{\nu}(T_{d})\,\nu^{\beta}$) and related to the
Rayleigh-Jeans spectral index by $\beta\,=\,\alpha_{FIR}\,-\,2$). The perceived
magnitude of this excess is critically dependent on the quality of the
observations and the model extrapolations.  The Magellanic Cloud SEDs
are much better sampled, and the WMAP observations extend the spectral
coverage to millimeter wavelengths, obviating the need for an
extrapolation. This leads to a significant improvement in the quality
of fits, and thus the determination of excess emission.

The LMC and the SMC exhibit a striking excess at both millimeter and
sub-millimeter wavelengths.  It is noteworthy that equally
well-sampled spectra of the Orion complex in the 1-1000 GHz range
(Dicker et al. 2009) show a similar range of turnover frequencies. In
the relatively quiet HII region the upturn frequency is at about 100
GHz -- as it is in the star-burst galaxies. However, towards the
star-forming Orion-KL source, the upturn frequency occurs at 40 GHz,
as it does in the LMC.  Simultaneously, the dust emissivity changes
from $\beta\,=\,2.0$ in the nebula to $\beta\,=\,1.2$ in Orion-KL.

The excess emission manifests itself between 0.3 mm and 10 mm (30 GHz
and 1000 GHz) as a combination of a relatively low upturn frequency
(defined as the frequency of minimum emission in the millimeter
spectral range, $\nu_{mm}^{min}$) {\it and} a relatively low
sub-millimeter spectral index $\alpha_{FIR}$, i.e. a relatively low
emissivity $\beta$ around unity.  There are two obvious ways in which
the dust emission upturn will shift to lower frequencies (longer
wavelengths) in any spectrum without the need to invoke special
properties of the radiating dust.  First, if the radio continuum is
unusually {\it weak} with respect to the far-infrared emission from
dust, the downshift of the radio part will quite naturally cause the
point where dust becomes dominant (the upturn) to shift to lower
frequencies.  This is not the case as the Magellanic Clouds have, in
fact, a relatively strong radio continuum.  The comparison galaxies
from Sect. 2.4 (included in Fig.\,\ref{galwmap} have (much) larger
${S_{max}^{dust}}\over{S_{min}^{mm}}$ ratios (2000-3000) than the
Magellanic Clouds (1650 and 1200) but do not exhibit a lower upturn
frequency.  Second, if the same far-infrared emission peak occurs at a
lower frequency, the Rayleigh-Jeans tail is displaced to lower
frequencies by the same amount, also causing the upturn to occur
earlier.  In the Magellanic Cloud, the far-infrared peak indeed occurs
at somewhat longer wavelengths $\lambda_{dust}^{max}$ (lower
frequencies) than in most other galaxies but not by the amount needed
to explain the observed low upturn frequencies $\nu_{mm}^{min}$.

We refer to Fig. 6 of Leroy et al. (2007) to illustrate that the SMC
spectrum is much flatter than the canonical $\beta$ = 2 spectrum. It
is even in excess of the $\beta$ = 1.5 model spectrum, and the
longest-wavelength data point at 1200 $\mu$m suggests further
flattening.  The same situation applies to the LMC spectrum.  In
addition to the unmistakeable sub-millimeter excess of the LMC and the
SMC, the data presented in this paper show {\it a further millimeter
  excess}.

We have quantified this additional millimeter excess by calculating at
four frequencies the combined emission of the non-thermal radio,
thermal radio, and thermal dust components. For the former, we
extrapolated the results from the preceding discussion; for the latter
we extrapolated the spectral fits to the DIRBE and TopHat data
published by Aguirre et al. (2003) in their Fig. 4 and Table 9,
assuming no additional excess to be present at their lowest frequency of
245 GHz (longest wavelength of 1200 $\mu$m).  Our results are given
in Table\,\ref{Magexcess}, which shows a significant excess (about
50$\%$ of the total emission at 93 GHz) in the LMC, and an even
stronger excess in the SMC at all mm wavelengths.

\subsubsection{Possible explanations for a (sub)millimeter excess}

Various dust emission mechanisms have been suggested in order to
explain excess emission at millimeter and shorter wavelengths.

(i) {\it Very cold big dust grains.} Any far-infrared/sub-millimeter
dust emission spectrum can be modelled by a sufficient number of
modified black-body curves each representing a population of big grain
dust particles in thermal equilibrium at a particular dust
temperature.  An emissivity $\beta\,=\,2$ is commonly assumed,
although an inverse temperature dependence
$\beta\,=\,(0.40\,+\,0.008\,T_{d})^{-1}$ has been proposed by Dupac et
al. (2003).  The values $\beta\,\approx\,1.7$ characterising the
star-burst disk galaxies correspond to mean dust temperatures
$T_{d}\,\approx\,25$ K.  However, any such fit of the FIR/sub-mm
spectra of the Magellanic Clouds requires the additional presence of a
significant component of very cold dust ($T_{d}\,<<\,10 K$,
approaching 3 K).  At such low temperatures, very large amounts of
cold dust are required to produce even a modest amount of excess
emission.  Regarding other galaxies for which a sub-millimeter excess
was surmised, virtually all (Lisenfeld et al. 2002; Dumke et al. 2004;
Bendo et al. 2006; Zhu et al. 2009) authors have rejected this
solution as they consider the implied great masses of cold dust and
the resulting low gas-to-dust ratios to be implausible, in addition to
the difficulty of finding large amounts of very cold dust precisely in
those environments where low metallicities provide the least shielding
against strong ambient radiation fields.  These arguments apply here
as well.  The SMC has a lower metallicity than the LMC, yet it shows a
higher excess. It is hard to imagine how it could be richer in cold
dust.

A more detailed look at the distribution of WMAP emission over the
Clouds (Fig.\,\ref{localspec}) suggests the same conclusion.  There is
more excess emission from the bright star-forming regions 30 Doradus,
N~11, N~44 than from the LMC Bar and the quiescent northern edge near
N~48, although the field centered on the moderately bright southern
region N~206 exhibits the highest excess. In the SMC, the NE and SW
Bar regions are of very different appearance but show practically
identical spectra; the SMC Wing has the steepest spectrum.  It is hard
to see how these patterns could correspond to the distribution of very
cold dust in either galaxy.

(ii) {\it Different dust grain composition or structure.}  The
Magellanic Clouds may host a population of dust grains with optical
properties very different from those of the dust grains in more
metal-rich galaxies.  Structurally different dust grains, such as
fluffy or fractal particles (Ossenkopf $\&$ Henning 1994; Paradis et
al. 2009) could produce the sub-millimeter excess. In fact, Reach et
al. (1995) propose this to be the explanation for the widespread cold
dust component they found in the Milky Way from COBE/FIRAS
measurements, after rejecting either very cold big dust grains or very
small grains as a possibility.

The dust grain emissivity $\beta$ is not only a function of dust
temperature, but it also depends on the grain composition. For
instance, amorphous graphite has $\beta\,=\,1$ whereas crystalline
dust has $\beta\,=\,2$ (Mennella et al. 1998; Agladze et al. 2005).
Thus, we cannot exclude in low-metallicity star-forming galaxies such
as the LMC and the SMC the predominance of a dust grain population
with different optical properties causing $\beta$ to be around unity,
or even the inconspicuous presence of such dust in more massive
metal-rich spiral galaxies (such as the Milky Way, cf. Reach et
al. 1995). However, it is not clear what this population is (but see
M\'eny et al. 2007), or what processes would lie at the root of its
dominance in both star-forming dwarf galaxies and more extreme
(ultra)luminous infrared galaxies, nor is it obvious that $\beta$
could be as low as zero, as appears to be the case in the SMC.

(iii) {\it Very small spinning dust grains}, first modelled by Draine
$\&$ Lazarian (1998), have been invoked to explain anomalously high
and apparently dust-correlated microwave emission in the WMAP Milky
Way foreground, and Murphy et al. (2009) have suggested that such
grains are also present in the nearby spiral galaxy NGC~6946.
Recently, Ali-Ha\"{i}moud al. (2009) and Dobler et al. (2009) have
suggested that this peak frequency may occur anywhere between 30 GHz
and 50 GHz, but the spectra in Fig.\,\ref{Magspec} lack a maximum
around these frequencies. However, the WIM presented by these authors
may not be representative of what is expected in the Magellanic
Clouds, as they calculated spinning dust emissivities for grains
illuminated by an ISRF with intensity U = 1 (Mathis et al. 1983).  In
actual fact, the IR modelling of the SMC and LMC SEDs requires a
distribution of radiation field intensities from U = 0.1-0.8 to
1000. This may shift the peak to much higher frequencies up to 100 GHz
(Ysard $\&$ Verstraete 2009, and Ysard et al. 2009).  The possibility
that the LMC and SMC excess involves spinning dust cannot be excluded,
and is explored in more detail in a companion paper (Bot et al. 2010)

The presence of the excess emission in the Magellanic Cloud
sub-millimeter spectrum has an important consequence. As long as its
nature is unidentified, it is impossible to ascertain the degree to
which it `contaminates' the Rayleigh Jeans tail of the far-infrared
dust emission. {\it It follows that fitting existing dust emission models
to this tail will not provide information on the amount of dust much
cooler than that responsible for the far-infrared emission peak,
i.e. on dust radiating at temperatures below $\approx15-20$ K.  As
even large amounts of cold dust contribute only modest amounts of
emission, the apparently indeterminate nature of the Rayleigh-Jeans
tail renders reliable determination of total dust mass and gas-to-dust
ratio impossible}.

The Magellanic Clouds (and other star-forming dwarf irregular
galaxies) differ in a number of ways from more massive galaxies, such
as the Milky Way, M~82, NGC~253, in their dust-related properties.  In
addition to the pronounced millimeter and sub-millimeter excess
emission and the weaker 12$\mu$m emission, they have (a) lower
metallicities, (b) fewer PAHs, (c) much lower total extinction, (d)
much weaker $\lambda$2175\AA\, extinction features, and (e) more
steeply rising UV extinction curves.  In {\it all} these respects, the
SMC is more extreme than the LMC, and they appear to be related in
some way.  A common denominator may be metallicity-related
modifications of individual dust grains or the global dust population,
or both.

\section{Summary and conclusions}

\begin{enumerate}
\item
We have extracted from the COBE-DIRBE and WMAP databases maps of the
Large and the Small Magellanic Cloud in the 1.25 $\mu$m - 240 $\mu$m
and 23 GHz - 93 GHz spectral ranges respectively.  We have used the
maps to determine globally integrated flux densities.
\item
We complemented the COBE-DIRBE and WMAP flux densities by those
literature flux densities that reliably represent the global emission
from the Clouds. We used the resulting data sets to construct the flux
density and energy distributions over the full spectral range from
low-frequency radio to ultraviolet, for the first time covering the
critical three spectral decades in the sub-millimeter-to-centimeter
window (10 GHz - 1 THz).
\item
We have established that the SMC and the LMC have significant emission
above the expected free-free radio continuum starting at frequencies
of 10 GHz - 30 GHz and extending over millimeter and sub-millimeter
wavelengths into the far-infrared.  
\item
The excess is not caused by cold, big dust grains.  The existence of
the excess emission will provide new insight in the nature of
interstellar dust, but in the meantime renders impossible reliable
determination of total dust mass as well as gas-to-dust ratio. 
\item
The free-free thermal radio continuum is $13.4\,(\nu/10GHz)^{-0.1}$ Jy
for the SMC, and $146\,(\nu/10GHz)^{-0.1}$ Jy for the LMC, implying
Lyman continuum fluxes log\,$N_{L}\,=\,51.72$ and
log\,$N_{L}\,=\,52.57$, respectively.
\item
The mean visual extinctions internal to the SMC and the LMC are
$A_{V}^{int}\,=\,0.45$ mag and $A_{V}^{int}\,=\,0.56$ mag
respectively, in addition to Milky Way foreground extinctions of 0.12
mag and 0.25 mag.
\end{enumerate}

\newpage

\appendix

\section{Spectral data used}

%Table 1  WMAP Data

\begin{table*}
\caption[]{Large Magellanic Cloud}
\begin{center}
\begin{tabular}{ccccl}
\hline
\noalign{\smallskip}
Freq.     & Wavel.       & \multicolumn{2}{c}{Flux Densities} \\
 $\nu$   & $\lambda$  &  Jy & $\Delta$Jy  & Reference\\
(GHz)    & (mm)          &\multicolumn{3}{c}{}\\
\noalign{\smallskip}
\hline
\noalign{\smallskip}
0.0197  &  15230       & 5270   &  1054	   & Shain$^{a}$ 1959 \\
0.045   &   6670       & 2997   &  450     & Alvarez et al. 1987 \\
0.0855  &   3510       & 3689   & 400	   & Mills 1959$^{a}$ \\
0.0968  &   3100       & 2839   & 600	   & Mills 1959$^{a}$	\\ 
0.158   &   1900       & 1736   & 490	   & Mills 1959$^{a}$ \\
0.408   &    735       & 925    & 30	   & Klein et al. 1989 \\
0.408   &    735       & 934    & 173      & Haslam et al. 1981$^{a}$ \\ 
1.4     &    214       & 529    & 30	   & Klein et al. 1989 \\
1.4     &    214       & 450    & 30	   & Hughes et al. 2006	\\ 
2.3     &    130       & 404    & 30	   & Klein et al. 1989 \\
2.3     &    130       & 412    & 50       & Mountfort et al. 1987 \\
2.45    &    122       & 390    & 20       & Haynes et al. 1991 \\
4.75    &     63       & 363    & 30       & Haynes et al. 1991 \\
4.75    &     63       & 296    & 25	   & Hughes et al. 2006 \\
8.55    &     35       & 270    & 35       & Haynes et al. 1991 \\
23	&     13       & 160	& 25	   & This Paper 2010 \\
33	&     9.1      & 159	& 25	   & This Paper 2010 \\
41	&     7.3      & 165	& 25	   & This Paper 2010 \\
61      &     4.9      & 205	& 32	   & This Paper 2010 \\
93	&     3.2      & 366	& 65	   & This Paper 2010 \\
245	&     1.2      & 1630   & 170	   & Aguirre et al. 2003 \\
400	&    0.75      & 7930   & 590	   & Aguirre et al. 2003 \\ 
460	&    0.65      & 10570  &  890	   & Aguirre et al. 2003 \\
630	&    0.48      & 29660  & 2980	   & Aguirre et al. 2003 \\
1250    &    0.24      & 112300 & 2400	   & Aguirre et al. 2003 \\
1250    &    0.24      & 153000 & 3000 	   & This Paper 2010 \\
2140    &    0.14      & 177700 & 3800	   & Aguirre et al. 2003 \\
2140    &    0.14      & 251100 & 4500 	   & This Paper 2010 \\	
3000    &    0.10      & 130100 & 17600    & Aguirre et al. 2003 \\\
3000    &    0.10      & 200200 & 3000	   & This Paper 2010 \\
3000    &    0.10      & 206250 & 30000    & Hughes et al. 2006 \\
3000    &    0.10      & 184687 & 28000	   & Rice et al. 1988 \\
5000    &    0.06      & 82917  & 12000	   & Rice et al. 1988 \\
5000    &    0.06      & 50500 	& 7500     & This Paper 2010 \\
5000    &    0.06      & 96250  & 14000    & Hughes et al. 2006 \\
12000   &    0.025     & 7824   & 1200	   & Rice et al 1988 \\
12000   &    0.025     & 7520  	& 1100	   & This Paper 2010 \\
25000   &    0.012     & 2782   & 400	   & Rice 1988 \\
25000   &    0.012     & 3450   & 600      & This Paper 2010 \\
61220   &   0.0049     & 1240	& 190      & This Paper 2010 \\
85710	&   0.0035     & 2190	& 300      & This Paper 2010 \\
136365  &   0.0022     & 3765	& 400      & This Paper 2010 \\
240000  &   0.00125    & 4520	& 650      & This Paper 2010 \\	
542000	&   0.00055    & 2520   & 139      & de Vaucouleurs et al. 1991 \\
681000	&   0.00044    & 1840	&  87      & de Vaucouleurs et al. 1991 \\
819000	&   0.000365   & 783	&  53      & de Vaucouleurs et al. 1991 \\
2142850 &   0.00014    & 204	&  40      & Page \& Carruthers 1981 \\
2307700 &   0.00013    & 228	&  45      & Page \& Carruthers 1981 \\
\noalign{\smallskip}
\hline
\noalign{\smallskip}
\end{tabular}
\end{center}
Note: $^{a}$ As revised and listed by Klein  et al. 1989
\end{table*}

\begin{table*}
\caption[]{Small Magellanic Cloud}
\begin{center}
\begin{tabular}{ccccl}
\hline
\noalign{\smallskip}
Freq.     & Wavel.       & \multicolumn{2}{c}{Flux Densities} \\
 $\nu$    & $\lambda$    &  Jy & $\Delta$Jy  & Reference\\
(GHz)     & (mm)         & \multicolumn{3}{c}{}\\
\noalign{\smallskip}
\hline
\noalign{\smallskip}
0.0197    & 15230        &5270   & 1054 & Shain 1959; cf Alvarez et al. 1989\\
0.045     & 6670         & 415   &  80  & Alvarez et al 1989 \\
0.0855    & 3510         & 460   & 200  & Mills 1959$^{a}$ \\
0.408     & 735          & 133   &  10  & Haslam 1982$^{a}$ \\
0.843     & 356          & 75    &  8   & Ye and Turtle 1991 \\
1.4 	  & 214          & 42    &  6   & Loiseau et al. 1987$^{b}$ \\
2.3       & 130          & 31    &  6   & Mountfort et al. 1987 \\
2.45      & 122          & 26    &  3   & Haynes et al 1991 \\
4.75      & 63           & 19    &  4   & Haynes et al 1991 \\
8.55      & 35           & 15    &  4   & Haynes et al 1991 \\
23	  & 13           & 17    &  3   & This Paper 2010 \\
33	  & 9.1          & 20    &  4   & This Paper 2010 \\
41	  & 7.3          & 24    &  5   & This Paper 2010 \\
61	  & 4.9          & 40    &  8   & This Paper 2010 \\
93	  & 3.2          & 88    &  18  & This Paper 2010 \\         
245	  & 1.2          & 320   &  80  & Aguirre et al. 2003 \\
400	  & 0.75         & 950   & 190  & Aguirre et al. 2003 \\
460	  & 0.65         & 1620  & 290  & Aguirre et al. 2003 \\
630	  & 0.48         & 3200  & 810  & Aguirre et al. 2003 \\
1250	  & 0.24         & 12070 & 540  & Aguirre et al. 2003 \\
1250	  & 0.24         & 12350 & 1440 & This Paper 2010 \\
1250	  & 0.24         & 9600  & 4400 & Stanimirovic et al. 2000 \\
1765      & 0.17  1      & 18500 & 4000 & Wilke et al 2004$^{c}$ \\
1875	  & 0.16         & 19000 & 4300 & Leroy et al.2007 \\
2140	  & 0.14         & 20300 & 920  & Aguirre et al. 2003 \\
2140	  & 0.14         & 18900 & 2005 & This Paper 2010 \\
2140	  & 0.14         & 14000 & 5600 & Stanimirovic et al. 2000 \\
3000	  & 0.10         & 16480 & 2220 & Aguirre et al. 2003 \\
3000	  & 0.10         & 15750 & 2125 & This Paper 2010 \\
3000	  & 0.10         & 13600 & 930  & Stanimirovic et al. 2000 \\
3000	  & 0.10         & 13900 & 1810 & Schwering 1988 \\
4285	  & 0.07         & 12500 & 2400 & Leroy et al.2007 \\
5000	  & 0.06         & 8450  & 730  & This Paper \\
5000	  & 0.06         & 6700  & 1060 & Stanimirovic et al. 2000 \\
5000	  & 0.06         & 7170  & 957  & Schwering 1988 \\
12500	  & 0.024        & 350   & 50   & Leroy et al. 2007 \\
12000	  & 0.025        & 460   & 180  & Stanimirovic et al. 2000 \\
12000	  & 0.025        & 425   & 70   & This Paper \\
12000	  & 0.025        & 310   & 170  & Stanimirovic et al. 2000 \\
12000	  & 0.025        & 385   & 56   & Schwering 1988 \\
25000	  & 0.012        & 80    & 30   & Stanimirovic et al. 2000 \\
25000	  & 0.012        & 105   & 30   & Stanimirovic et al. 2000 \\
25000	  & 0.012        & 147   & 26   & Stanimirovic et al. 2000 \\
25000     & 0.012        & 235   & 25   & This Paper 2010 \\
37500	  & 0.008        & 78    & 8    & Bolatto et al. 2007 \\
51274	  & 0.0058       & 85    & 9    & Bolatto et al. 2007 \\
61220	  & 0.0049       & 110   & 43   & Stanimirovic et al. 2000 \\
61220     & 0.0049       & 150   & 25   & This Paper 2010 \\
66667	  & 0.0045       & 115   & 12   & Bolatto et al. 2007 \\
83333	  & 0.0036       & 163   & 16   & Bolatto et al. 2007 \\
85710	  & 0.0035       & 220   & 31   & Stanimirovic et al. 2000 \\
85710	  & 0.0035       & 280   & 40   & This Paper 2010 \\
136365	  & 0.0022       & 400   & 37   & Stanimirovic et al. 2000 \\
136365	  & 0.0022       & 525   & 65   & This Paper 2010 \\
240000	  & 0.00125      & 510   & 25   & Stanimirovic et al. 2000 \\
240000	  & 0.00125      & 670   & 80   & This Paper 2010 \\
542000	  & 0.00055      & 458   & 46   & de Vaucouleurs et al. 1991 \\
681000	  & 0.00044      & 354   & 34   & de Vaucouleurs et al. 1991 \\
819000    & 0.000365     & 181   & 22   & de Vaucouleurs et al. 1991 \\
\noalign{\smallskip}
\hline
\noalign{\smallskip}
\end{tabular}
\end{center}
Notes: 
$^{a}$ As revised and listed by Loiseau et al. (1987); 
$^{b}$ As revised and listed by Haynes et al. (1991); 
$^{c}$ As revised by Leroy et al. (2007)
\end{table*}

\end{document}